\documentclass{jfm}
\usepackage{color}
\usepackage{graphicx}
\usepackage{natbib}
\usepackage{psfrag,epsfig}
\usepackage{amsmath,rotating, subfig}
\usepackage{setspace}
\usepackage{amssymb}
\usepackage{stmaryrd}
\usepackage{tensor}
\usepackage{rotating}
\usepackage{hyperref}
\usepackage{cleveref}
\usepackage{listings}
\usepackage{extarrows}
\usepackage{multirow}
\ifCUPmtlplainloaded \else
  \checkfont{eurm10}
  \iffontfound
    \IfFileExists{upmath.sty}
      {\typeout{^^JFound AMS Euler Roman fonts on the system,
                   using the 'upmath' package.^^J}%
       \usepackage{upmath}}
      {\typeout{^^JFound AMS Euler Roman fonts on the system, but you
                   dont seem to have the}%
       \typeout{'upmath' package installed. JFM.cls can take advantage
                 of these fonts,^^Jif you use 'upmath' package.^^J}%
      }
  \else
  \fi
\fi

\ifCUPmtlplainloaded \else
  \checkfont{msam10}
  \iffontfound
    \IfFileExists{amssymb.sty}
      {\typeout{^^JFound AMS Symbol fonts on the system, using the
                'amssymb' package.^^J}%
       \usepackage{amssymb}%
       \let\le=\leqslant  
         
      }{}
  \fi
\fi

\ifCUPmtlplainloaded \else
  \IfFileExists{amsbsy.sty}
    {\typeout{^^JFound the 'amsbsy' package on the system, using it.^^J}%
     \usepackage{amsbsy}}
    {}
\fi

\newsavebox{\astrutbox}
\sbox{\astrutbox}{\rule[-5pt]{0pt}{20pt}}

\usepackage{color,ulem}
\usepackage[usenames,dvipsnames,svgnames,table]{xcolor}
\definecolor{light-gray}{gray}{0.5}
\newcommand{\red}[1]{\textcolor{black}{#1}}
\newcommand{\cut}[1]{\textcolor{light-gray}{\sout{}}}

\newcommand{\Ma}{\text{M}}
\newcommand{\heater}{\text{f}}
\newcommand{\pade}{\alpha}

\newcommand{\leftS}{\text{\tiny{L}}}
\newcommand{\rightS}{\text{\tiny{R}}}

\title[Thermoacoustic instability -- a dynamical system and time domain analysis]
{Thermoacoustic instability -- a dynamical system and time domain analysis}

\author[T. Sayadi, V. Le Chenadec, P.J. Schmid, F. Richecoeur and M. Massot]%
{T\ls A\ls R\ls A\ls N\ls E\ls H\ns S\ls A\ls Y\ls A\ls D\ls
  I$^1$\thanks{Email address for correspondence:
    sayadi@ladhyx.polytechnique.fr},\ns
  V\ls I\ls N\ls C\ls E\ls N\ls T\ns L\ls E\ns C\ls H\ls E\ls N\ls A\ls D\ls E\ls C$^{2,3}$\break
  P\ls E\ls T\ls E\ls R\ns J.\ns S\ls C\ls H\ls M\ls I\ls D$^{1}$, \ns
  F\ls R\ls A\ls N\ls C\ls K\ns R\ls I\ls C\ls H\ls E\ls C\ls O\ls E\ls U\ls R$^{2,3}$\break
  \and M\ls A\ls R\ls C\ns M\ls A\ls S\ls S\ls O\ls T$^{2,3,4}$}

\affiliation{$^1$Department of Mathematics, Imperial College London, London SW7 2AZ, United Kingdom\\[\affilskip]
  $^2$CNRS - UPR 288, 92295, Laboratoire EM2C, Ch\^atenay-Malabry, France\\[\affilskip]
  $^3$\'Ecole Centrale Paris, 92295 Ch\^atenay-Malabry, France\\[\affilskip]
  $^4$ F\'ed\'eration de Math\'ematiques de l'\'Ecole Centrale Paris, CNRS - FR 3487, 92295 Ch\^atenay-Malabry, France}

\pubyear{xxxx}
\volume{xxx}
\pagerange{xxx--xxx}
\date{?; revised ?; accepted ?. - To be entered by editorial office}
\begin{document}

\maketitle

\begin{abstract}
This study focuses on the Rijke tube problem, which includes features relevant to the modeling of thermoacoustic coupling in reactive flows: a compact acoustic source, an empirical model for the heat source, and nonlinearities. This thermo-acoustic system features both linear and nonlinear flow regimes with complex dynamical behavior. In order to synthesize accurate time-series, we tackle this problem from a numerical point-of-view, and start by proposing a dedicated solver designed for dealing with the underlying stiffness, in particular, the retarded time and the discontinuity at the location of the heat source. Stability analysis is performed on the limit of low-amplitude disturbances by means of the projection method proposed by \citet{jarlebring08}, which alleviates the linearization with respect to the retarded time. The results are then compared to the analytical solution of the undamped system, and to Galerkin projection methods commonly used in this setting. This analysis provides insight into the consequences of the various assumptions and simplifications that justify the use of Galerkin expansions based on the eigenmodes of the unheated resonator. We illustrate that due to the presence of a discontinuity in the spatial domain, the eigenmodes in the heated case, predicted by using Galerkin expansion, show spurious oscillations resulting from the Gibbs phenomenon. Finally, time-series in the fully nonlinear regime, where a limit cycle is established, are analyzed and dominant modes are extracted. By comparing the modes of the linear to that of the nonlinear regime, we are able to illustrate the mean-flow modulation and frequency switching, which appear as the nonlinearities become significant and ultimately affect the form of the limit cycle. The analysis of the saturated limit cycles shows the presence of higher frequency modes, which are linearly stable but become significant through nonlinear growth of the signal. This bimodal effect is not captured when the coupling between different frequencies is not accounted for. In conclusion, a dedicated solver for capturing thermoacoustic instability is proposed and methods for analyzing linear and nonlinear region of the resulting time-series are introduced.
\end{abstract}




\section{Introduction}

Energy conversion devices typically involve combustion processes in a confined geometry, where various complex phenomena, related to fluid flow, acoustic radiation and thermal effects, interact. These interactions and couplings may act as sources of organized motions referred to as thermoacoustic instabilities. Examples of mechanisms driving such motions include oscillations in the supply of reactants, structural vibrations of the chamber, or burning processes. In the latter case, an unsteady flow interacting with a fluctuating thermal source, potentially caused by turbulent combustion, can result in an unsteady heat release. The unsteady heat release can in turn act as an acoustic source in the system resulting in self-sustained oscillations~\citep{rayleigh1896}. In the unstable case, the system usually reaches a limit cycle during which the pressure can oscillate considerably. As a result, thermoacoustic instabilities can affect the life and the performances of these technical devices. In addition, ever more stringent environmental regulations on the emissions of these devices tend to shift operating conditions towards the lean-burn parameter range, which is more prone to thermoacoustic instabilities. This trend has fueled the design and study of models able to predict the stability properties of these self-sustained oscillations.

For modeling purposes, the zone of heat release is considered as compact. This assumption introduces some inaccuracies in the predicted frequencies, but it is usually found to yield good predictions since the modes of interest are of low frequencies~\citep{dowling95}. The corresponding wavelengths are long compared with the duct diameter, and the problem may thus be considered one-dimensional. Additionally, the hydrodynamic region associated with the heat source is short compared with the length of these dominant acoustic wavelengths, and can be accounted for either using asymptotic analysis~\citep{mariappansujith2011} or as a discontinuity at a single axial location. The unsteady heat release can then be linked to the acoustic waves either by an analytical expression (\citet{dowling97} for flames; \citet{Heckl90} for an electrical heater) or by experimental measurements of the transfer function~\citep{Schuller03}. The eigenmodes and eigenfrequencies can then be determined by a dispersion relation in the frequency domain. This linear approach has recently been extended to weakly non-linear stability analysis by measuring the describing function of the heat source, and interesting results have been obtained in the prediction of limit cycles~\citep{Boudy2013}. These spectral techniques however forbid the prediction of complex transients due to frequency coupling, known to influence the bifurcation diagrams of the system. 

In this work, therefore, we investigate the evolution of the velocity and pressure oscillations from the small-amplitude linear stage to the fully nonlinear regime and provide time-series comparable to those from experimental observations. This objective is pursued with a minimum of additional assumptions and approximations, stemming from physical modeling or a choice of numerical techniques. The most commonly used methods in performing a nonlinear analysis for these types of instabilities are based on Galerkin expansions.  In this approach a system of partial differential equations is reduced to a set of coupled ordinary differential equations with time-varying coefficients by a projection onto a given basis~\citep{culick88}. The growth rates and amplitudes are estimated by an energy balance across the heat source. In these methods, typically, terms up to second or third order are retained. A particularly common choice of basis functions for the Galerkin expansion consists of the acoustic eigenfunctions of the duct without the heating element. This method has been applied in the Rijke tube setting (\citealt{sujith08}, \citealt{subramanian10}, \citealt{juniper10}, \citealt{sujith13}) to study the linear and nonlinear behavior of acoustic waves in the duct. It should be noted that this approach approximates a discontinuous spatial signal by continuous basis functions. The consequences of this assumption or choice will be further investigated in this study. Another method for computing nonlinear time-series for thermoacoustic systems has been proposed by \citet{dowling97}; the wave expansion technique has been employed to investigate nonlinear oscillations of a ducted flame. In this approach, the acoustic wave propagation is computed explicitly. However, the boundary conditions have to be known analytically to determine the linear wave propagation away from the source; in addition, damping across the geometry cannot be easily incorporated.

Experimental investigations have shown that thermo-acoustic instabilities involve a limited number of modes related to the geometry of the system, suggesting that the full dynamics can be reproduced with low order models. However the complexity of the mode interactions in the transient phase requires tools able to compute and describe these couplings. In this study, we propose and advocate a dedicated solver for computing nonlinear time-series for the acoustic behavior of the aforementioned simplified system. The solver is flexible in the sense that boundary conditions can be easily modified to include appropriate reflections, measured experimentally. Moreover, any damping due to the interaction with the wall boundary layers of the duct or caused at the end of the duct can be accounted for. The choice of the heating model allows for variety, as long as it satisfies the compactness criterion. \red{The aim is to develop a solver for performing acoustic integration so that it could later be easily coupled with the relevant hydrodynamic calculation of the heat source.} Most importantly, this approach enables a time-series analysis without the need for assuming a shape for the eigenfunctions of the system {\it{a priori.}}  Consequently, linear stability analysis can be performed on the high-fidelity system of equations, and the eigenvalues and eigenfunctions of the linear problem can be computed directly, without resorting to further simplifying assumptions. We have adopted the projection method of~\citet{jarlebring08} to extract these eigenfunctions in the most general way. Dynamic mode decomposition \citep{schmid2010} is applied to the fully nonlinear regime to extract spatial modes of dynamical significance from the limit-cycle behavior. As a result, the modes extracted in the linear and nonlinear regimes can be compared directly, which will provide insight into the system dynamics throughout the evolution of the time-series.

This paper is organized as follows. In \S~\ref{sec:GE} the equations governing the acoustic propagation of waves inside a tube are described, and a model for the heating source is introduced. \S~\ref{sec:NM} describes the proposed dedicated solver, designed to accurately capture the interaction of the acoustic waves and the heat source. In addition, further simplifying assumptions --- leading to models which have been studied previously in an attempt to capture and quantify these instabilities --- are presented in \S~\ref{sec:models}. For low-amplitude disturbances these different approaches for analyzing the Rijke tube problem are critically assessed and compared in \S~\ref{sec:CDM}. The high-fidelity approach, based on our dedicated solver, is then applied to the linear and nonlinear regimes, and the dynamical behavior of the system is analyzed in both limits in \S~\ref{sec:fullanalysis}. Finally, a summary of our results and concluding remarks are presented in \S~\ref{sec:conclusion}.

\section{Governing equations}
\label{sec:GE}
\begin{figure}
  \centering
  {\includegraphics[width = 0.7\textwidth] {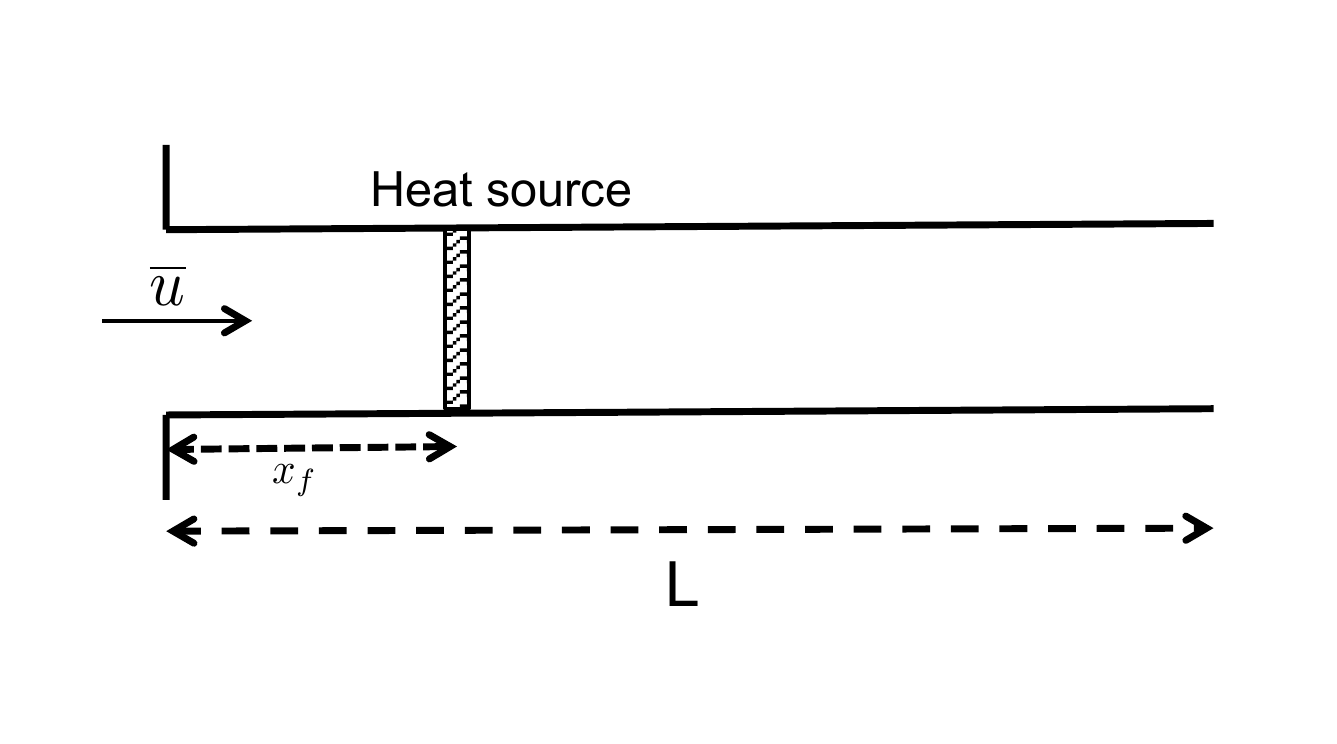}}
  \caption{Schematic of a horizontal duct with a compact heat source.}
  \label{fig:CD}
\end{figure}

A horizontal duct with a compact heat source is a convenient framework for studying the fundamental principles of thermoacoustic instabilities. A schematic of the tube is shown in figure~\ref{fig:CD}.
Since in this type of geometry longitudinal modes are most relevant, we can consider the equations of motion in a one-dimensional setting. Moreover, the boundary layers forming at the sides of the tube are much thinner than the diameter of the tube, $\delta_{99} \ll D$, with $\delta_{99}$ as the boundary layer thickness. Therefore, we take the inviscid equations to describe the acoustic propagation of waves outside the boundary layers and neglect their growth. This approach has been extensively used in the current setting~\cite{Williams1994,dowling95}, and a brief derivation is included here. The thermodynamic variables of the medium are expressed in terms of the pressure and entropy according to
\begin{equation}
  \frac{\text{D} \rho}{\text{D} t} =
  \left . \frac{\partial \rho}{\partial p} \right \vert_s
  \frac{\text{D} p}{\text{D} t} + \left . \frac{\partial \rho}{\partial s}
  \right \vert_p \frac{\text{D} s}{\text{D} t}.
\end{equation}
In the absence of diffusive effects, the first law simply states $\rho T \text{D} s / \text{D} t = q$, where $q$ denotes the heat release per unit volume and unit time. Finally, for a perfect gas with
\begin{equation}
  \left . \frac{\partial \rho}{\partial s} \right \vert_p =
  - \frac{\rho T \left ( \gamma -1 \right )}{c^2},
\end{equation}
we may use the continuity equation to write
\begin{equation}
  \frac{\text{D} p}{\text{D} t} + \rho c^2
  \frac{\partial u}{\partial x} = \left ( \gamma-1 \right ) q.
\end{equation}
The momentum equation simply reads,
\begin{equation}
  \frac{\text{D} u}{\text{D} t} + \frac{1}{\rho} \frac{\partial p}{\partial x} = 0.
\end{equation}
The linearization of these two equations about a homogeneous mean state, together with an acoustic scaling of the time variable, yields the following system of equations
\begin{equation}
  \left \{ \begin{aligned}
      \frac{\partial u^\star}{\partial t^\star} + \text{M}
      \frac{\partial u^\star}{\partial x^\star} + \frac{1}{\gamma \text{M}}
      \frac{\partial p^\star}{\partial x^\star} & = 0, \\
      \frac{\partial p^\star}{\partial t^\star} + \text{M}
      \frac{\partial p^\star}{\partial x^\star} + \gamma \text{M}
      \frac{\partial u^\star}{\partial x^\star} & = \gamma \text{M} Q^\star,
    \end{aligned} \right .
\end{equation}
where the identity $\rho c^2 = \gamma p,$ valid for a perfect gas, has been employed; in addition, we have
\begin{equation}
  Q^\star \equiv \frac{\left ( \gamma -1 \right ) L q'}{\overline{\rho}
    \overline{c}^2 \overline{u}}.
\end{equation}
The variables $\left ( p^\star, u^\star \right )$ denote the fluctuating pressure and velocity of the system, nondimensionalized by the mean velocity $\overline{u}$ and pressure $\overline{p},$ respectively. Furthermore, as is typical in this simplified setting, we neglect the effects of mean convection.

An additional modeling assumption consists in incorporating dissipative phenomena into the system. Dissipation may be introduced locally to model the finite hydrodynamic region in the vicinity of the heater~\citep{Heckl2007}, or globally to include end losses and viscous effects associated with the boundary layer, for example. Although the growth of the latter has been neglected, we nonetheless include a frequency-dependent term in the energy equation, finally leading to (the suffixes ${}^\star$ are omitted)
\begin{equation}
  \left \{ \begin{aligned}
      \frac{\partial u}{\partial t} + \frac{1}{\gamma \text{M}}
      \frac{\partial p}{\partial x} & = 0, \\
      \frac{\partial p}{\partial t} + \gamma \text{M}
      \frac{\partial u}{\partial x} + \xi \ast p & = \gamma \text{M} Q.
  \end{aligned} \right .
  \label{eq:GE}
\end{equation}
(*) is the convolution operator (due to frequency dependent damping). The frequency dependent damping accounts for acoustic losses due to imperfect reflection at the tube ends, as well as for boundary layer losses, and is prescribed by the following equation~\citep{howe98}
\begin{equation}
\xi_j = \frac{1}{\pi} \frac{\omega_j^2 A}{\overline{c} \, L} + \frac{1}{\sqrt{2}} \frac{P \sqrt{\omega}_j}
{A} \left ( \sqrt{\nu} + \left ( \gamma - 1 \right ) \sqrt{\lambda} \right ),
\end{equation}
where $\left ( P, \, L \right )$ denote the perimeter and the length of the tube, $\left ( \nu, \, \lambda \right )$ the kinematic viscosity and thermal diffusivity. It is difficult to account for this term in usual wave-expansion techniques given the incompatibility with the traveling wave solution. In simplified form, the damping is presented according to~\citet{Matveev2003a} and \citet{Sterling1991} by
\begin{equation}
\xi_j = \left ( c_1 j^2 + c_2\sqrt{j} \right ).
\label{eq:damping}
\end{equation}
This damping has previously been applied by~\citet{sujith13} and~\citet{juniper10} to model thermoacoustic instabilities in identical settings. The parameters $\left ( c_1, c_2 \right )$ in~(\ref{eq:damping}) are kept constant for all modes and are set to $c_1 = 0.1$ and $c_2 = 0.06$. These values are chosen to resemble experimental conditions and are conventionally used when employing the Rijke tube model \citep{sujith13}. Using the above definition for damping, the system of equations (\ref{eq:GE}) is then closed by providing a relation between the fluctuating heat release and the flow perturbations $\left( u, p \right )$. One such closure is provided below in the context of the Rijke tube.

\subsection{Modeling of the heat source}
Thermoacoustic instabilities are characterized by an interplay of multiple complex phenomena. A systematic understanding of the underlying dynamics and instability mechanisms can be gained from studying the simplified model of the Rijke tube. This model offers a convenient setup and testbed for both numerical and experimental investigations. The original Rijke tube~\citep{rijke1859} consisted of a vertical tube, in which a mean flow arose due to natural convection. A horizontal tube, in contrast, offered a more controlled setup, where an electrical gauze replaced the flame as a localized heat source (\citealt{Heckl90}, \citealt{matveev02}). Mathematical models of the heat source in the tube have been reviewed by~\citet{raun93}. The model adopted in this study is the one commonly used in previous numerical approaches (\citealt{sujith13}, \citealt{juniper10}). The model used for the heat release in this study has been proposed by \cite{sujith08} in studying non-modal effects in thermoacoustic systems. This nonlinear model employs King's law \citep{king14}, which describes the heat transfer of a cylindrical hot-wire in a perpendicular flow as
\begin{equation}
  \overline{Q} + Q' = L_w(T_w -T)\left[ \kappa + 2\sqrt{\pi \kappa c_v
  \bar{\rho}\frac{d}{2}|u|}\right],
\end{equation}
where $L_w$ and $d$ are the length and the diameter of the wire, $T_w$ is the wire temperature, $T$ denotes the temperature of the surrounding air, $\kappa$ stands for the heat conductivity of the air and $c_v$ represents the specific heat of the air per unit mass at constant volume. There is a time lag (delay) $\tau$ between the heat transfer and the velocity as a result of thermal inertia. \cite{lighthill54} showed that $\tau \approx 0.2\frac{d}{\bar{u}},$ for Reynolds numbers greater than $10$ and frequencies small compared to $20\bar{u}/d$.

King's law predicts the nonlinearity in the heat release, for velocity amplitudes greater than the mean velocity, when flow reversal occurs at the heat source. This relation did not agree with experimental observations of~\cite{Heckl90}, where it was found that nonlinear behavior commences at velocity fluctuations about one-third of the mean velocity. Therefore, the above equation has been modified to reflect this observation, resulting in
\begin{equation}
  \overline{Q} + Q'  = L_w(T_w -T)\left[ \kappa + 2\sqrt{\pi \kappa c_v
  \bar{\rho}\frac{d}{2}}\left((1-\dfrac{1}{3\sqrt{3}})\sqrt{\bar{u}}
  + \frac{1}{\sqrt{3}} \sqrt{\left \vert  \frac{\bar{u}}{3} 
  + u'(t-\tau)\right \vert}\right)\right]. 
  \label{eq:HR}
\end{equation}
Nondimensionalizing~(\ref{eq:HR}), as described previously, and subtracting the mean, we obtain the system of partial differential equations~\ref{eq:GE} with
\begin{equation}
  Q = Q_\heater(t-\tau)\delta(x-x_\heater) = \dfrac{K}{2}
  \left[\sqrt{|1/3 + u_\heater(t-\tau)|} - \sqrt{1/3}\right]\delta(x-x_\heater),
  \label{eq:heater}
\end{equation}
and $\delta$ denoting the standard Dirac distribution. Usually, $u_\heater$ is taken as the velocity value on the cold side ($x \le x_\heater$) of the heater. In this manner, the system of equations is closed; it provides the starting point for our numerical discretization.

\section{Numerical and analytical solution of the governing equations}
\label{sec:NM}
The spatial discretization scheme for computing high-fidelity solutions of the governing equations is briefly described in the following section. In addition, the projection method used for approximating the eigenvalues and eigenvectors of the delay differential equations in the limit of small-amplitude disturbances is introduced. Finally, an analytical solution for the undamped linear system is derived. The method employed for the instability analysis, together with the analytical inviscid solution, provide the basis for a comparison of alternative models, presented in \S~\ref{sec:models}, and for an assessment of various levels of numerical approximations.

\subsection{High-fidelity numerical solution}
\label{ss:SDS}
In order to provide a numerical approximation of the continuous system, we proceed by discretizing the tube with a staggered arrangement of the variables. The velocity nodes are located at the cell centers, and the pressure nodes at the cell faces. This arrangement is particularly suited for the open-end boundary conditions used here (Dirichlet boundary conditions for the pressure, Neumann boundary conditions for the velocity). A compact Pad\'e approximation is employed for the spatial derivatives in order to limit dispersion errors, which would be detrimental to the propagation of the acoustic waves. A Ghost-Fluid treatment~\citep{Fedkiw1999} of the velocity discontinuity across the heating element is employed to retain the compact character of the acoustic source. Furthermore, to achieve first-order accuracy across the singularity, the heater location is chosen to coincide with a pressure node. The space discretization is completed by the approximation of the damping, performed by one-sided discrete Fourier transforms in order to alleviate Gibbs phenomenon-related artifacts. These approximations, presented in this section, finally lead to a system of coupled ordinary differential equations with delay, which are then advanced in time using a high-order time integrator dedicated to delay-differential equations~\citep{guglielmi01}.

The compactness of the nonlinear term, in the spatial discretization scheme, is preserved by recasting the second equation in~(\ref{eq:GE}) as
\begin{equation}
  \dfrac{\partial p}{\partial t}  + \gamma \Ma \,
  \dfrac{\partial}{\partial x} \left [ u - Q_\heater \left ( t-\tau \right ) H
  \left ( x - x_\heater \right ) \right ]  + \xi \ast p = 0.
\end{equation}
In the above, $H$ stands for the Heaviside function, which shows that the velocity field contains a discontinuity of amplitude $Q_\heater\left ( t-\tau \right )$ at the location of the heater. As mentioned above, for enhanced accuracy we enforce the heater location to coincide with a cell face, denoted by the index $i_\heater$ and the abscissa $x_\heater = x_{i_\heater}.$ Furthermore, we determine the magnitude of the velocity term in $Q_\heater$ as
\begin{equation}
    u_\heater \left ( t - \tau \right ) \simeq \frac{U_{i_\heater-1/2}
    \left ( t-\tau \right ) +U_{i_\heater+1/2} \left ( t-\tau \right )}{2}.
    \label{eq:disc_heater_strength}
\end{equation}
This definition is used to permit a fair and detailed comparison with the Galerkin methods presented in \S~\ref{sec:models}, which assume a continuous velocity profile. \red{Furthermore, for the purpose of comparison with the Galerkin-based approaches, the change in the mean state is also not considered.}

We introduce the following notations: an approximation of the velocity (resp., pressure) gradient, denoted by $G_{i}$ (resp., $H_{i-1/2}$), is required at the cell faces (resp., centers).  For all $i \in \left\llbracket 1, N-1\right \rrbracket$, the velocity gradients are approximated by means of the following compact Pad\'e scheme~\citep{nag03},
\begin{equation}
  \pade G_{i-1} + \left ( 1-2 \pade\right ) G_i + \pade G_{i+1} =
  \frac{U_{i+1/2}-U_{i-1/2}}{\Delta_i} -
  \frac{Q_\heater \left ( t-\tau \right )}{\Delta_{i}} \delta^{i_\heater}_i +
  \mathcal{O} \left ( \Delta_i^p \right ),
\end{equation}
where $\delta_i^{i_\heater}$ denotes the Kronecker delta symbol. By setting $\pade = 1/24,$ this approximation achieves $4^{\text{th}}$-order ($p=4$) accuracy, even at the boundaries where homogeneous Neumann boundary conditions are provided at $X_0$ and $X_{N}.$ One exception is the heater cell boundary, where the order of accuracy is dependent on the Ghost-Fluid approximation. For this particular cell, it can be shown that the first-order extrapolation of the jump, contained in the above formula, yields a first-order accurate discretization of the velocity gradient, where the lowest-order term of the truncation error is proportional to
\begin{equation}
  \left . \frac{\partial^2 u}{\partial x^2} \right \vert_{X_{i_\heater}^+} -
  \left . \frac{\partial^2 u}{\partial x^2}
  \right \vert_{X_{i_\heater}^-}.
\end{equation}
The approximation of the pressure gradient for $i \in \left \llbracket1, N \right \rrbracket$ is given by
\begin{equation}
  \pade H_{i-3/2} + \left ( 1-2 \pade\right ) H_{i-1/2} +
  \pade H_{i+1/2} = \frac{P_{i}-P_{i-1}}{\Delta_{i-1/2}}
  + \mathcal{O} \left ( \Delta_{i-1/2}^p \right ),
\end{equation}
where $\pade = 1/24$ for the interior cells ($p=4$). For the boundary cells, the right-hand side accounts for homogeneous Dirichlet boundary conditions ($\pade$ is set to $0$), causing the accuracy to locally revert to second order ($p=2$). 

Although the pressure signal is continuous across the heater location, its derivative is not. Not explicitly accounting for this jump may result in the appearance of spurious damping as a consequence of Gibbs phenomenon. Although only first order, the effect may easily be alleviated by performing discrete Fourier transforms on either side of the tube, thus avoiding the convolution across the discontinuity~\citep{Canuto}.

To summarize, the order of accuracy of the spatial scheme is nominally fourth-order according to the Pad\'e scheme inside the duct, but second-order at the boundaries and first-order across the discontinuity (heating element). The fully coupled, nonlinear and discretized system of equations is then solved using the time integrator RADAR5, developed by~\citet{guglielmi01}. This dedicated  software relies on high-order error estimates and time-step predictions, as well as the detection of breaking points, to yield accurate solutions of systems of delay-differential equations.

\subsection{Spectral and bifurcation analysis}
\label{ss:LA}

In the limit of small-amplitude disturbances, the nonlinear source term in~(\ref{eq:heater}) can be linearized as follows
\begin{equation}
  Q_\heater \left ( t-\tau \right ) \equiv
  \dfrac{K}{2}\left[\sqrt{|1/3 + u_\heater(t-\tau)|} -
  \sqrt{1/3}\right]
  \approx \sqrt{3} \dfrac{K}{4}u_\heater(t-\tau),
  \label{eq:LST}
\end{equation}
by simply keeping the first-order term of the Taylor series expansion. In this study, the stability analysis is performed by means of a projection method proposed by~\citet{jarlebring08}, which avoids the linearization of the retarded term. The analysis is performed on the discretized system of equations described in \S~\ref{ss:SDS}, with a modified source term given by (\ref{eq:LST}). The three parameters in the ensuing bifurcation analysis are the strength $K$ of the heater, its location $x_\heater$, and the delay $\tau.$ The semi-discretized system of delay-differential equations can be recast in matrix form as follows
\begin{equation}
  \dot{x} = A_0 x + A_1 x(t-\tau),
\end{equation}
where, in the discretized system of \S~\ref{ss:SDS}, we have $x =\{U_{1/2}, P_1, U_{3/2}, \cdots, P_{N-1}, U_{N-1/2}\}$, accounting for the staggered-grid arrangement of the governing variables. The matrix $A_0$ contains the differentiation and damping matrices, and $A_1$ includes the linearized source term. This\cut{ linear} system has eigenvalues and eigenvectors which satisfy the relation
\begin{equation}
  (-\lambda I + A_0 + A_1e^{-\tau \lambda}) v = 0. \label{eq:eig}
\end{equation}
In the above expression, $v$ stands for the eigenvectors and $\lambda$ for the eigenvalues of the above set of coupled delay-differential equations. This characteristic equation includes an exponential term, in contrast to standard linear eigenvalue problems. This term categorizes (\ref{eq:eig}) as a nonlinear eigenvalue problem. This exponential acts as a source for multiple extra roots of the transcendental characteristic equation, $\textrm{det}(-\lambda I + A_0 + A_1e^{-\tau \lambda}) = 0$, which in general cannot be computed or expressed analytically. Projection methods have been applied by~\citet{jarlebring08} to solve this nonlinear eigenvalue problem numerically. They are incorporated in the software package DDE-BIFTOOL~\citep{engelborghs00}. An alternative method for solving this nonlinear eigenvalue problem is introduced by \citet{selimefendigil2011}. For relatively large systems of equations, the projection method is found to be superior to other methods, for example, a discretization of the solution operators or an equivalent boundary-value problem formulation of the delay-differential equation. We have implemented the system of ordinary delay-differential equations described above, together with the simplified models given in \S~\ref{sec:models}, in this software package and will use the collocation method to find the numerically approximated eigenvalue spectra and the resulting bifurcation diagrams.

\subsection{Analytical solution of the undamped linear system}
\label{ss:AS}
In the absence of singularities, the undamped linear system admits traveling-wave solutions on either side of the heater ($\epsilon \in \left \{\text{L}, \text{R} \right \}$)
\begin{equation}
  \left \{ \begin{aligned}
      p_\epsilon & = A^+_\epsilon \exp \left (\mathsf{i} k x' -
        \mathsf{i} \omega t \right ) +
      A^-_\epsilon \exp \left ( - \mathsf{i} k x' -
        \mathsf{i} \omega t \right ), \\
      u_\epsilon & = A^+_\epsilon \exp \left ( \mathsf{i} k x' -
        \mathsf{i} \omega t \right ) -
      A^-_\epsilon \exp \left ( -\mathsf{i} k x' -
        \mathsf{i} \omega t \right ),
    \end{aligned} \right .
\end{equation}
where $x' = x - a$ ($a = x_\heater$) if $\epsilon = \text{R}$, and $x' = x$ otherwise. Based on the acoustic nondimensionalization presented previously, we have $\omega = k$. An analytical solution of this problem may be found for more general conditions (variable cross-sections, presence of a mean flow, etc). We focus here on the simplified Rijke-tube setup. For the idealized open-end boundary conditions adopted above, we relate the coefficients $A_{\leftS,\rightS}^\pm$ to the reflection factors
\begin{equation}
  \left \{ \begin{aligned}
      R_\leftS & = \frac{A_\leftS^+}{A_\leftS^-} = -1, \\
      R_\rightS & = \frac{A_\rightS^+}{A_\rightS^-} \exp \left
      ( 2 \mathsf{i} k b \right ) = -1,
    \end{aligned} \right.
\end{equation}
where, $a + b = 1$\cut{, with $L$ denoting the length of the tube}. Two more relations are obtained by matching the profiles across the heating element according to
\begin{equation}
  \begin{aligned}
    p_\rightS - p_\leftS & = 0, \\
    u_\rightS - u_\leftS & = \frac{\sqrt{3}}{4} K \left [ W u_\leftS
    \left ( x_f, t-\tau \right ) + \left ( 1-W \right ) u_\rightS
    \left ( x_f, t-\tau \right ) \right ],
  \end{aligned}
\end{equation}
with $W \in \left [ 0, 1 \right ]$. In the $n-\tau$ model~\citep{poinsot05}, we have $n=\sqrt{3} K / 4$ and $W$ is set to $1$. For the proposed discretization, both left and right velocity values are available; the parameter $W$ may be set to any value. In the case of the Galerkin projection introduced in \S~\ref{sec:models}, however, the velocity profile is assumed continuous. For a fair comparison of both discretizations, an average over the two one-sided limits ($W = 1/2$) is used. For a non-trivial solution to exist, $\omega$ must satisfy a dispersion relation of the type
\begin{multline}
  \left ( 1+\frac{\sqrt{3} K W}{4} \exp\left ( \mathsf{i} \omega \tau
  \right )\right )
  \cos \left [ k a \right ] \sin \left [ k b \right ] \\
  + \left ( 1-\frac{\sqrt{3} K \left ( 1-W \right )}{4} \exp \left (
  \mathsf{i} \omega \tau \right ) \right ) \sin \left [ k a \right
  ] \cos \left [ k b \right ] = 0.
\end{multline}
This relation may easily be inverted numerically to find the (complex) eigenvalues $\omega$ and the eigenvectors of the system, as illustrated in \S~\ref{sec:CDM}. Both methods introduced in this study for analyzing the linear, \S~\ref{ss:LA}, and nonlinear, \S~\ref{ss:SDS}, behavior of the signal are flexible in the number and the form of the delay parameters. In addition, it would be \red{straightforward} to account for the velocity signal at one side of the discontinuity, as is done in the conventional $n-\tau$ model. However, due to the restrictions of the Galerkin-based methods, introduced in the following sections, and arising from the continuous basis functions, which force the average of the two states to be considered, we have adopted the same formalism for the sake of fairness when comparing the results. 

\section{Further simplifying assumptions}
\label{sec:models}

An alternative way of simplifying the system of partial differential equations governing the dynamics of acoustic waves in a tube, is by means of a Galerkin projection. A Galerkin projection reduces the system of partial delay-differential equations to a set of ordinary delay-differential equations. In the following section we will describe such models, developed on the basis of Galerkin expansions and widely applied to analyze thermoacoustic instabilities in the Rijke tube (\citealt{sujith08}, \citealt{subramanian10}, \citealt{juniper10}, \citealt{sujith13}).

\subsection{Galerkin expansion}
\label{ss:FCG}

A conventional approach for analyzing the thermoacoustic problem in a Rijke tube setting is the use of Galerkin projection and modal expansion. In these methods the spatially and temporally varying pressure and velocity signals are expanded in terms of purely spatial basis functions that inherently satisfy the boundary conditions. The temporal dependence of the signal is accounted for by means of time-varying coefficients. The choice of these basis functions is not unique. In the Rijke tube problem, for example, since the eigenfunctions of the system are not known {\it{a priori}}, the eigenfunctions of the self-adjoint acoustic part of the linearized system are chosen as a substitute. The basis functions are thus continuous sine and cosine functions, describing the eigenfunctions of a horizontal tube in the absence of the heater. As a result, the acoustic pressure and velocity are written as
\begin{equation}
  u = \sum_{j=1}^N \cos(k_jx) U_j(t) , \qquad \textrm{and} \qquad
  p = \gamma M \sum_{j=1}^N \cos(k_jx) P_j(t)
\end{equation}
where $k_j = j\pi$ is the nondimensional wavelength of the $j^{\textrm{th}}$ duct mode.  These basis functions form a complete basis in the limit as $N \to \infty;$ however, in practice, a finite number of modes is considered. Substituting this expansion into~(\ref{eq:GE}) gives
\begin{subeqnarray}
  && \dot{U_j} - j\pi P_j = 0, \\
  && \dot{P_j} + j\pi U_j  + \xi_j  P_j=
  K \sin(j\pi x_f)\left[\sqrt{\left|\frac{1}{3} + \sum_{k=1}^N
  \cos(k\pi x_f)U_k(t-\tau)\right|} - \sqrt{\frac{1}{3}}\right].
\label{eq:FCG}
\end{subeqnarray}
This model will be referred to as the fully-coupled Galerkin model, since the coupling between different frequencies is accounted for in the source term on the right-hand side of~(\ref{eq:FCG}b).

\subsection{Decoupling the heat source from the acoustics}
\label{ss:DG}

If we further assume that the influence of higher modes can be neglected and a single-mode analysis of the Rijke tube accurately reproduces the correct dynamics of the full system, we can then simplify the right-hand-side of~(\ref{eq:FCG}) by removing the summation over all wavelengths \citep{sujith13}, resulting in the equation
\begin{equation}
  \dot{P_j} + j\pi U_j  + \xi_j   P_j=
  K \sin(j\pi x_f)\left[\sqrt{\left|\frac{1}{3} +
  \cos(j\pi x_f)U_j(t-\tau)\right|} - \sqrt{\frac{1}{3}}\right].
\end{equation}
As a result, each mode is decoupled from the others, and the equation governing the evolution of a respective mode solely depends on the mode itself. This model will be referred to as the decoupled Galerkin model in the following sections.

\subsection{Linearizing the delay}
\label{ss:LD}

The fully-coupled Galerkin model described in \S \ref{ss:FCG} can be further simplified by linearizing the delay, which was initially proposed by \citet{sujith08} \red{and later used in \citet{magri2013} and some calculations in \citet{juniper10}}. This linearization is valid only for the Galerkin modes for which $\tau \ll T_j$, where $T_j = 2/j$ is the period of the $j^{th}$ Galerkin mode. Under this assumption, the delayed velocity term can be written as $U_j(t-\tau) \approx U_j(t) - \tau dU_j(t)/dt$. By linearizing the delay, we convert the transcendental dispersion relation in $\lambda$, given in (\ref{eq:eig}), to a more familiar linear dispersion relation. At first glance, approximating the exponential component by the first two terms of the Taylor series might not appear as an oversimplification, but considering the oscillatory nature of the solution for complex $\lambda$, this assumption could be detrimental, even in the limit of small delays, since many zero-crossings due to the oscillations would be neglected. This issue will be further investigated in the next section. A summary of all methods analyzed in this study is presented in table~\ref{tab:table1}.
\begin{table}
  \begin{center}
    \begin{tabular}{lccc}
      \multirow{2}{*}{Numerical method} & Galerkin & Decoupled & Linearized  \\
      &  expansion & wave-numbers & retarded time \\
      & & & \\
      High-fidelity solution \S~\ref{ss:SDS} & no & no & no \\
      Fully-coupled Galerkin \S~\ref{ss:FCG} & yes & no & no \\
      Decoupled Galerkin \S~\ref{ss:DG} & yes & yes & no \\
      Galerkin with linearized delay \S~\ref{ss:LD}  & yes & no & yes \\
    \end{tabular}
  \end{center}
  \caption{Methods for analyzing thermoacoustic instabilities in the Rijke tube setting.}
  \label{tab:table1}
\end{table}

\section{Stability analysis in the limit of small-amplitude disturbances}
\label{sec:CDM}
In this section, the stability diagrams of different methods in the limit of small-amplitude disturbances are compared. In this limit, an analytical solution of the system with no damping can be derived, as demonstrated in \S~\ref{ss:AS}, and will then be used as a reference to validate each model. As a first step, the governing equations are linearized as described in \S \ref{ss:LA} and the results using different discretizations (Galerkin and high-fidelity) are compared to the analytical solution. The parametric stability diagrams and eigenvectors arising from the various models are computed with DDE-BIFTOOL~\citep{engelborghs00} and then compared. Finally, the predictions of the parametric stability diagram are verified using the signal computed by the time-integration of the high-fidelity discretization with the non-linear source term. This study quantifies the consequences of the various simplifications in the linear regime and examines the growth of the signal in the initial stage of development.

\subsection{Ability of methods to capture spectra and modes}
\label{ss:SandM}
\begin{figure}
  \centering
\subfloat[Symmetric model] {\includegraphics[width = 0.5\textwidth] {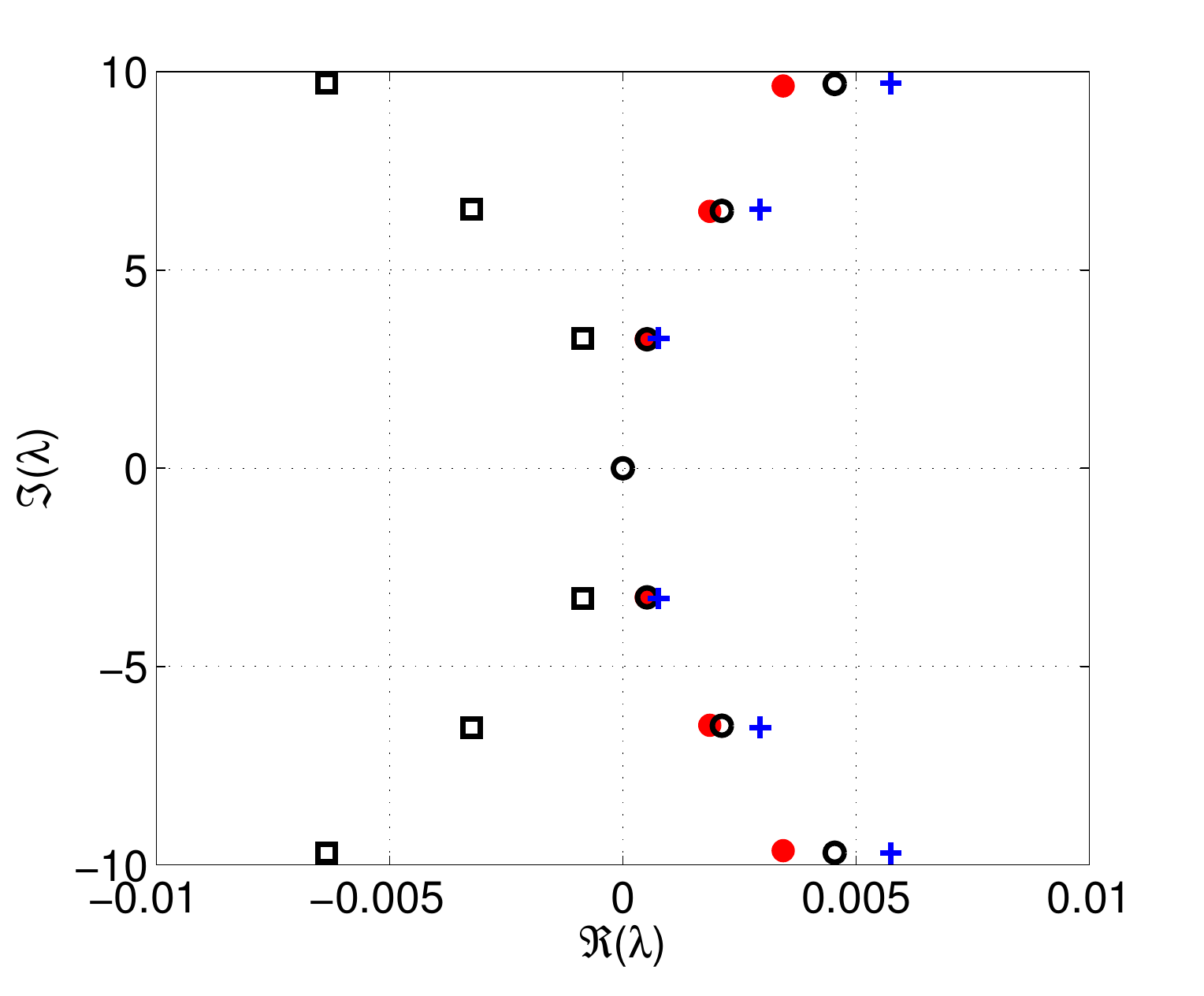}}
\subfloat[$n-\tau$ model] {\includegraphics[width = 0.5\textwidth] {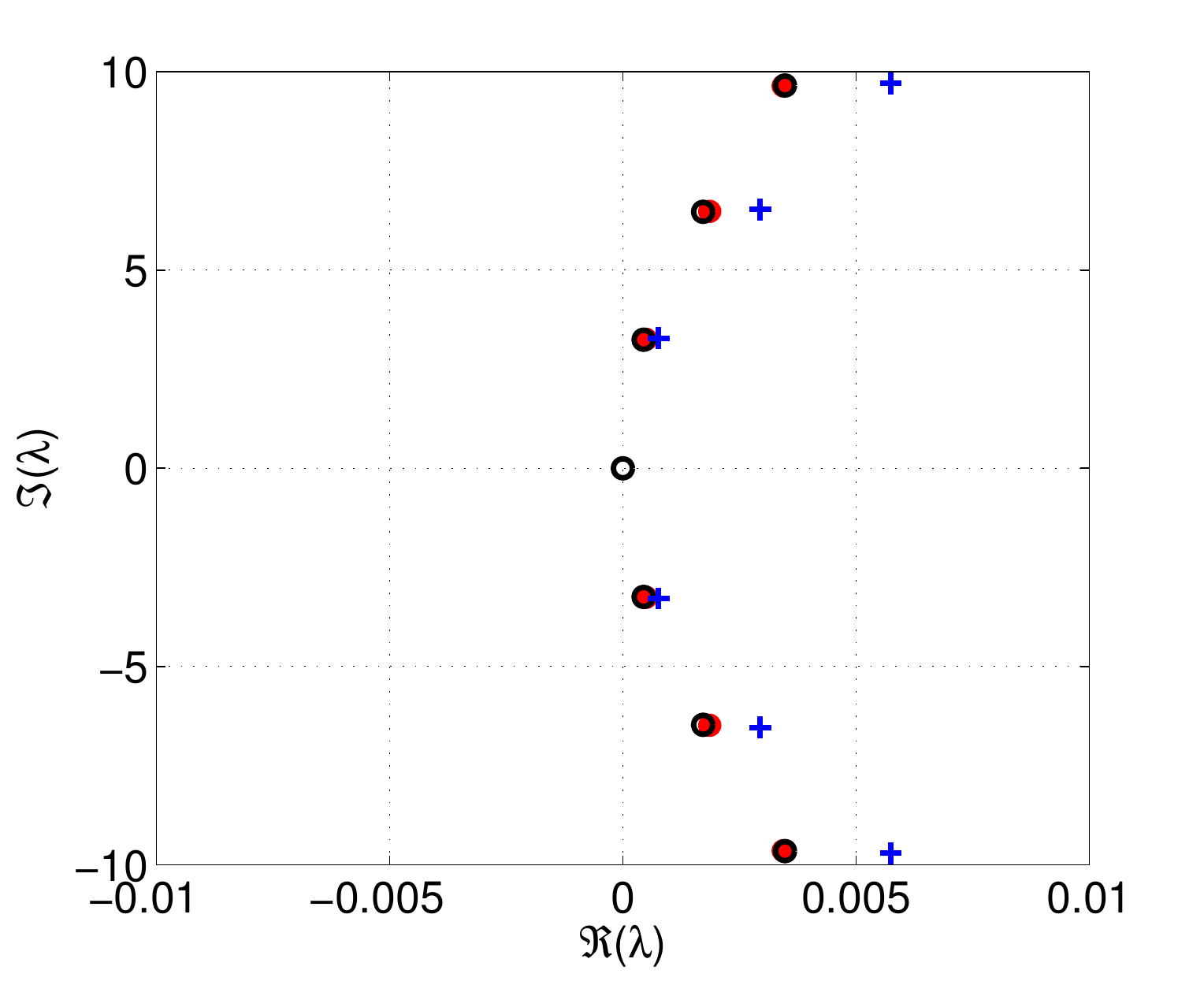}}
  \caption{Comparison of the analytical eigenvalue spectrum to that of the models. {\color{red}$\bullet$}, analytical (no damping); $\circ$,  high-fidelity system; $\square$, Galerkin with linearized delay; {\color{blue}$+$}, fully-coupled Galerkin. Parameters are given in the text.}
  \label{fig:ES}
\end{figure}

In the limit of small-amplitude disturbances and damping the models can be compared to the analytical solution of \S \ref{ss:AS}. The system parameters are chosen as $K = 1.3$, $x_f = 0.1$ and $\tau = 0.002$. The spectra of the linearized operators are compared in figure~\ref{fig:ES}, \red{and grid convergence studies have been performed, ensuring grid independence for all the results presented henceforth, by using a minimum of 96 grid points along the duct}. The delay has been chosen sufficiently small for a linearization of the delay (\S~\ref{ss:LD}) to be valid. In order to validate the predicted spectra against the analytical solution, the damping is set to a small value at $c_1 = \mathcal{O}(10^{-5})$ and $c_2 = \mathcal{O}(10^{-6})$, compared to the values specified in \S~\ref{sec:GE}. The decoupled Galerkin approach produces the same spectrum as the coupled Galerkin, as observed by \citet{sujith13}, and is therefore not included in figure~\ref{fig:ES} for the sake of clarity. Figure~\ref{fig:ES}(a) illustrates that linearizing the delay has the largest effect on the predicted eigenvalues. For example, the slightly unstable mode at frequency $\hbox{imag}(\lambda) \approx \pi$ appears as stable in the solution with the linearized delay, demonstrating that even the low frequencies are affected by this linearization. This effect causes a qualitative change in the stability behavior of the thermoacoustic system, suggesting a stable configuration at parameters where the analytical solution implies instability. The flexibility of the high-fidelity approach allows the direct comparison of the numerical spectrum to that of the $n-\tau$ solution, where the delay term is evaluated based on the left (upwind) state. The difference between the averaged and the upwind approximations is not accounted for by the Galerkin expansion. As seen in figure~\ref{fig:ES}(b), this leads to a deterioration of the Galerkin method performances. 

\begin{figure}
  \subfloat[ High-fidelity system]{\includegraphics[width = 0.53\textwidth] {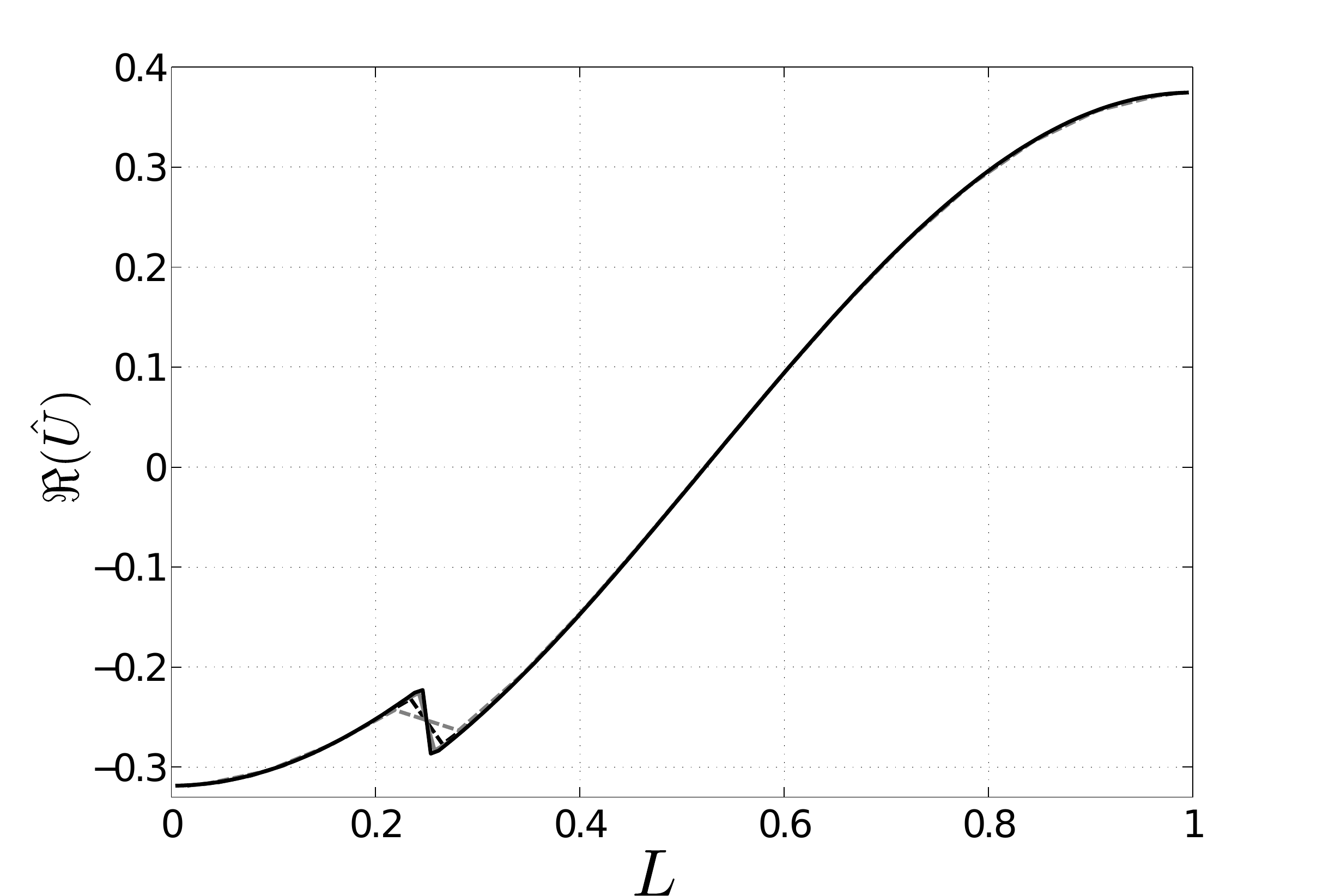}}
  \subfloat[ High-fidelity system, shock region]{\includegraphics[width = 0.53\textwidth] {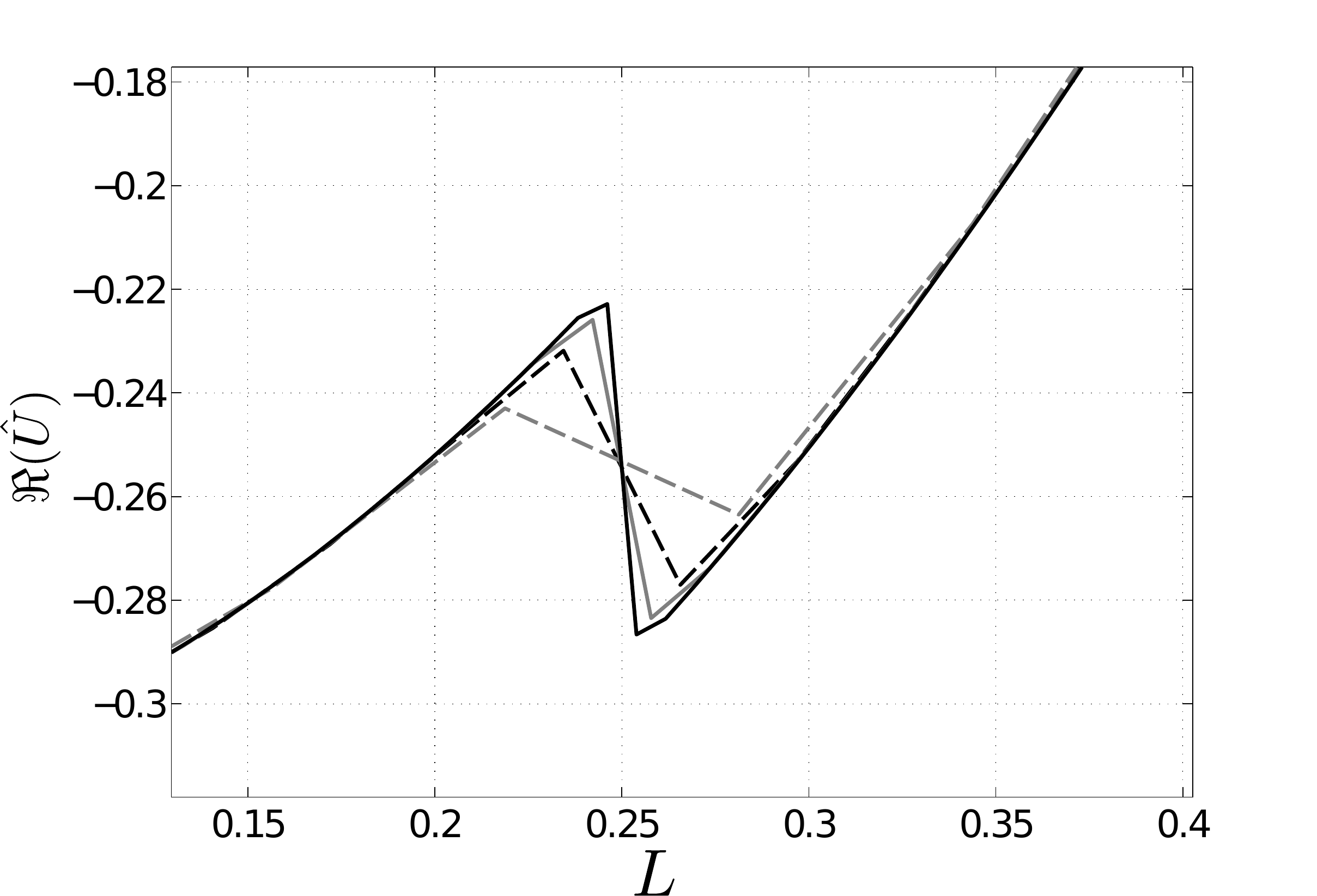}} \\
  \subfloat[ Fully Coupled Galerkin]{\includegraphics[width = 0.53\textwidth] {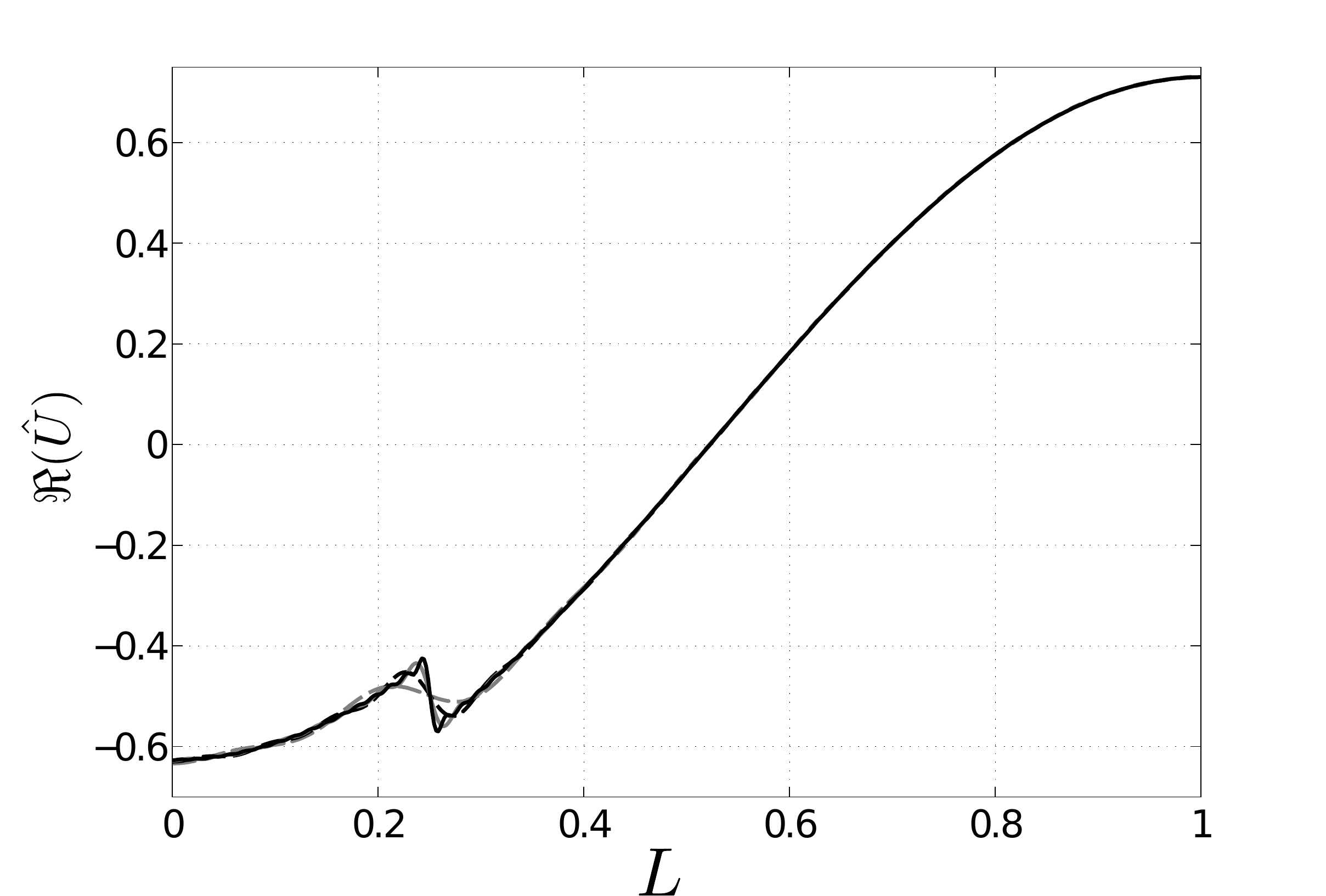}}
  \subfloat[ Fully Coupled Galerkin, shock region]{\includegraphics[width = 0.53\textwidth] {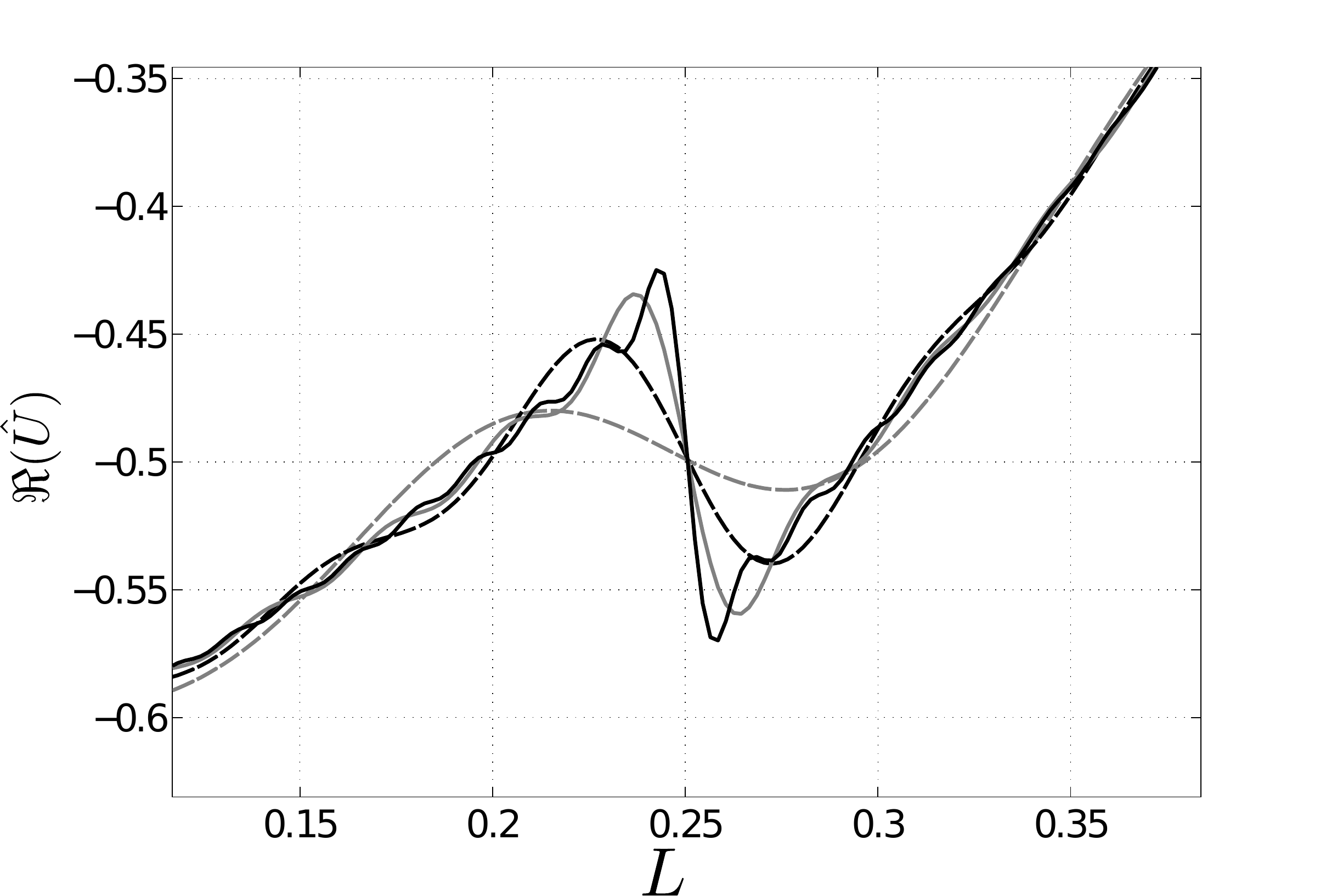}} \\
  \caption{Eigenvector of the velocity component of mode $\hbox{imag}(\lambda) = \pi$ for the linearized system ($x_f = 0.25$). Convergence for increasing the grid points/Galerkin modes (N).  $---$(Gray), $N = 16$; $---$, $N = 32$; -----(Gray), $N = 64$; -----, $N = 128$. }
  \label{fig:EIGVEC}
\end{figure}

\begin{figure}
  \centering
 \subfloat[Symmetric model]{\includegraphics[width = 0.5\textwidth] {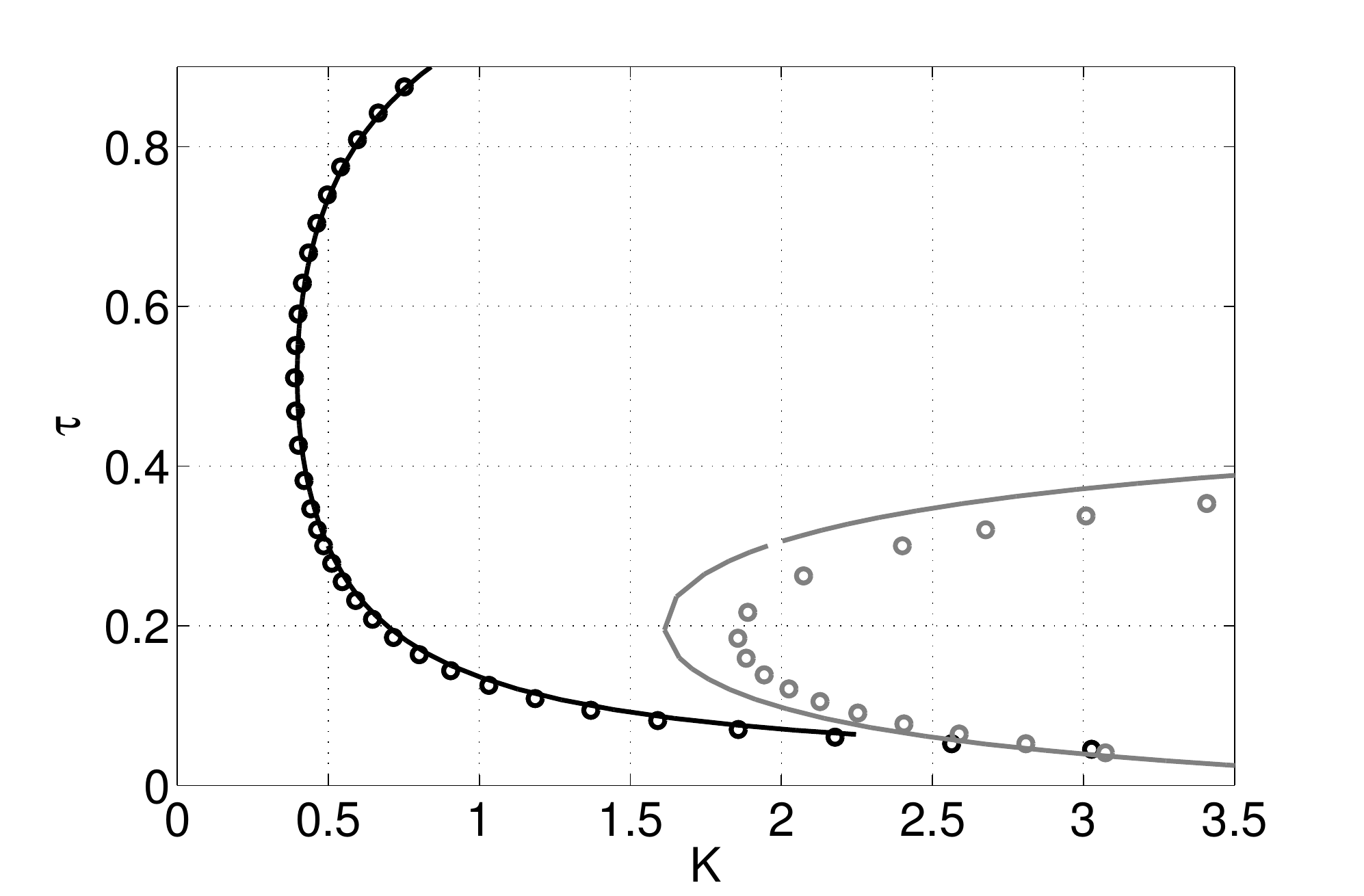}}
 \subfloat[$n-\tau$ model] {\includegraphics[width = 0.5\textwidth] {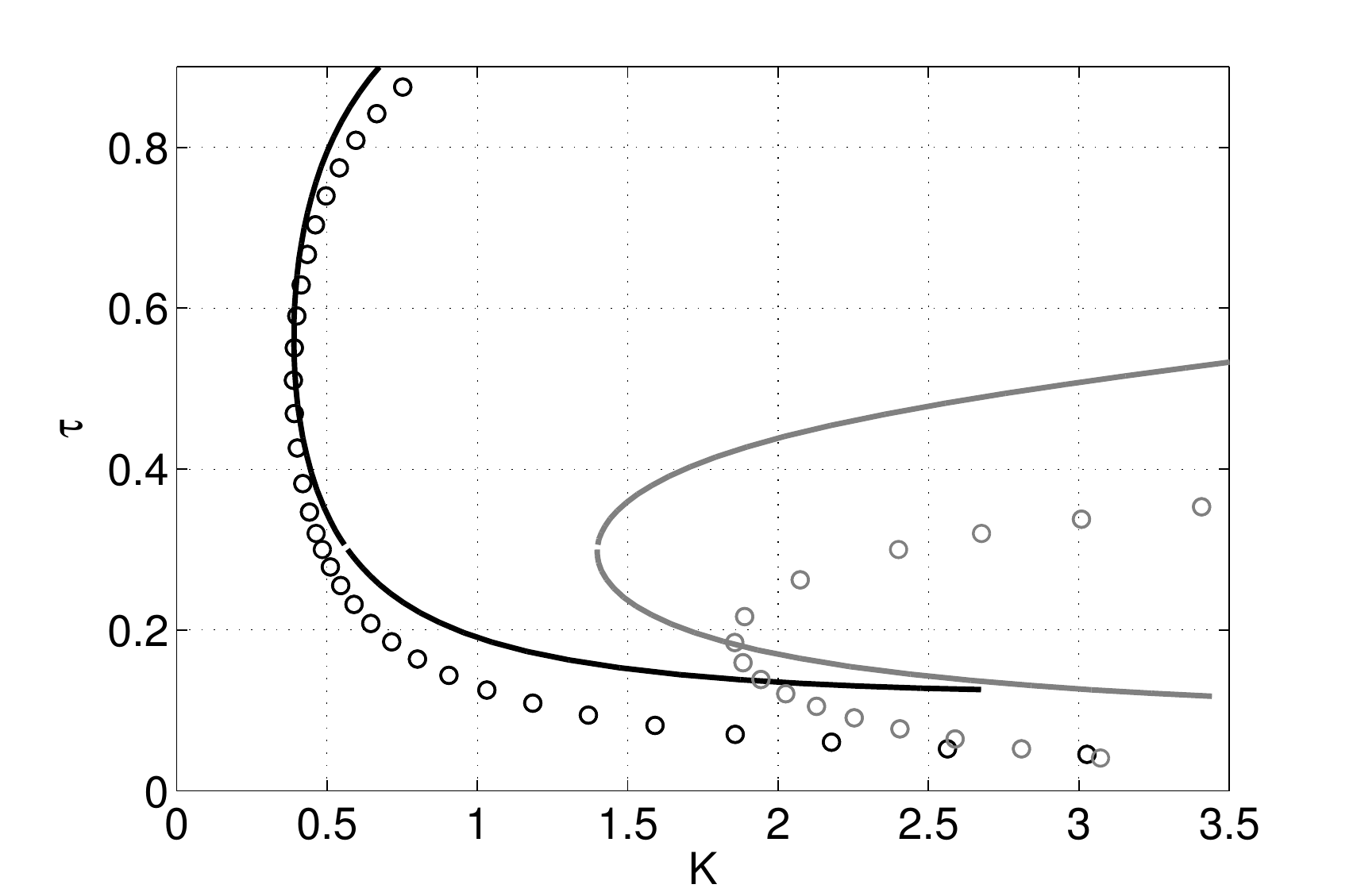}}
  \caption{Neutral curves of the first two frequency modes, for the highly damped solution, $x_f = 0.2$. $\hbox{imag}(\lambda) = \pi$: -----, high-fidelity; $\circ$, coupled Galerkin. $\hbox{imag}(\lambda) = 2\pi$: -----(Gray), high-fidelity; $\circ$(Gray), coupled Galerkin.}
  \label{fig:NC_lin}
\end{figure}

The differences between the Galerkin approach and the analytical solution can be explained by examining the respective eigenvectors computed using each model. Figure~\ref{fig:EIGVEC}(a) shows the convergence of the eigenvector corresponding to $\hbox{imag}(\lambda) \approx \pi$ as the number of grid points increases for the high-fidelity system. The behavior in the vicinity of the discontinuity is highlighted in figure~\ref{fig:EIGVEC}(b). This figure shows that the eigenvector is captured correctly. The same exercise is carried out for the fully-coupled Galerkin approach, the results of which are shown in figures~\ref{fig:EIGVEC}(c) and (d). The fully-coupled Galerkin expansion is in fact a spectral method applied to a solution with a fixed discontinuity. As a consequence, Gibbs phenomenon in the vicinity of the discontinuity is expected and observed. The Gibbs phenomenon results in higher-frequency oscillations with finite amplitudes, at the location of the discontinuity for each eigenvector. These oscillations become more relevant when systems with higher damping coefficients are considered. Following the governing equation, the damping becomes stronger as the wavelength of the oscillations decreases, therefore in such systems the presence of high frequency oscillations can ultimately result in over-damped solutions, as illustrated in figure~\ref{fig:NC_lin}(a). This figure compares the neutral curves of the high-fidelity and the coupled Galerkin approach for the lowest two frequencies, $\hbox{imag}(\lambda) = \pi$ and $2\pi$, of the duct, with the heater placed at a fixed position of $x_f = L/5$. The comparison shows that the neutral curve of the lowest frequency mode remains unchanged, regardless of the method. However, the envelope of instability for the mode with frequency $\hbox{imag}(\lambda) = 2\pi$ is underpredicted using the coupled Galerkin approach. For the choice of the averaged states the Galerkin expansion correctly evaluates the delayed source term, therefore this discrepancy can be attributed to the damping. For the $n-\tau$ model on the other hand, the inability of the Galerkin approach to account for an asymmetric source term leads to inaccurate prediction even at the lowest frequency, as seen in figure~\ref{fig:NC_lin}(b). This issue is avoided in the high-fidelity system, due to the special treatment of the discontinuity. As a result, the high-fidelity solution converges as the number of grid points increases (see figures \ref{fig:EIGVEC}(a) \& (b)) and approaches the analytical solution for small damping. The effect of high-frequency oscillations becomes clear in \S~\ref{sec:fullanalysis}, where the nonlinear limit cycles are studied.

\subsection{Instability diagram}
\begin{figure}
 \subfloat[$\hbox{imag}(\lambda) \approx \pi$]{\includegraphics[width = 0.33\textwidth] {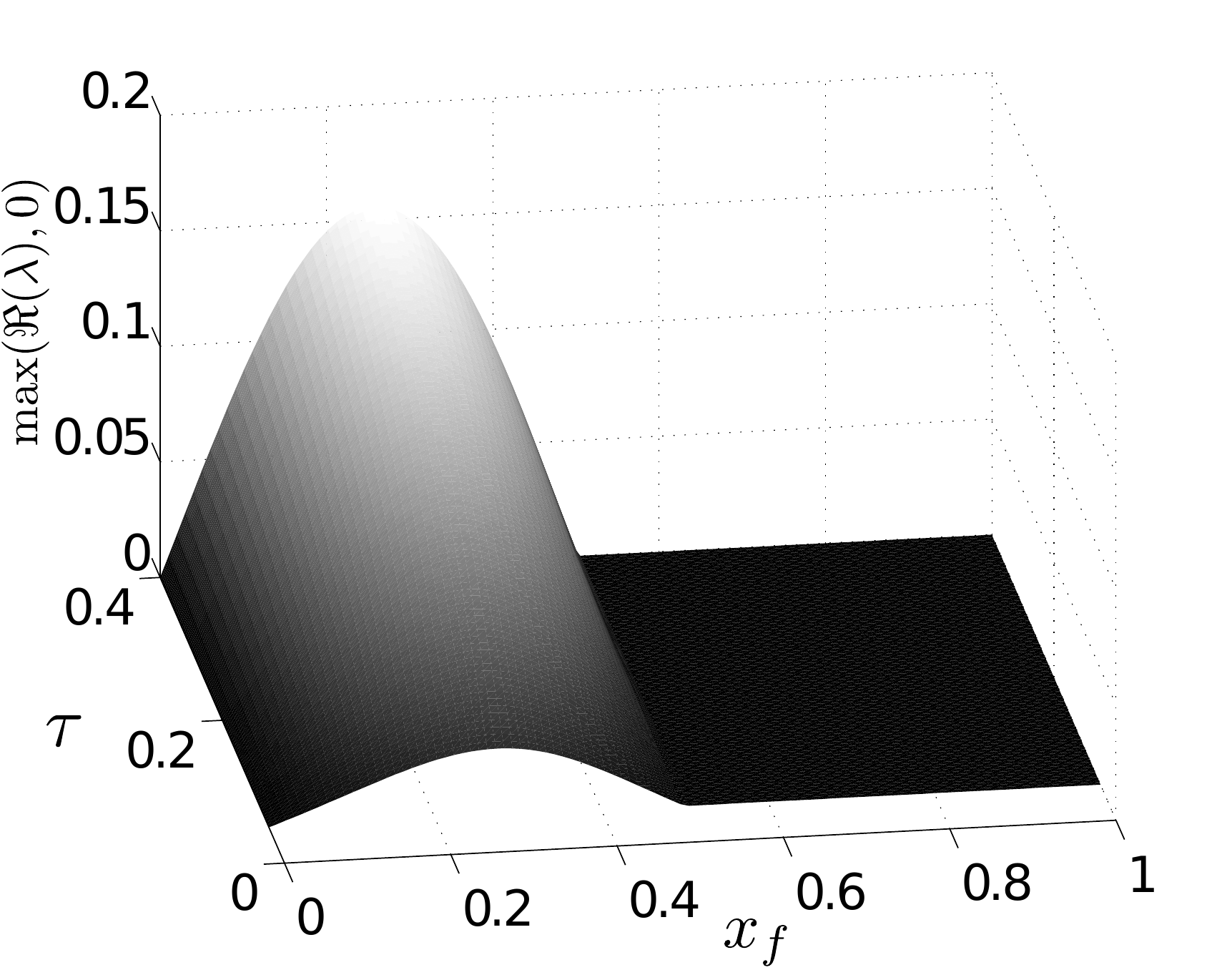}}
  \subfloat[$\hbox{imag}(\lambda) \approx 2\pi$]{\includegraphics[width = 0.33\textwidth] {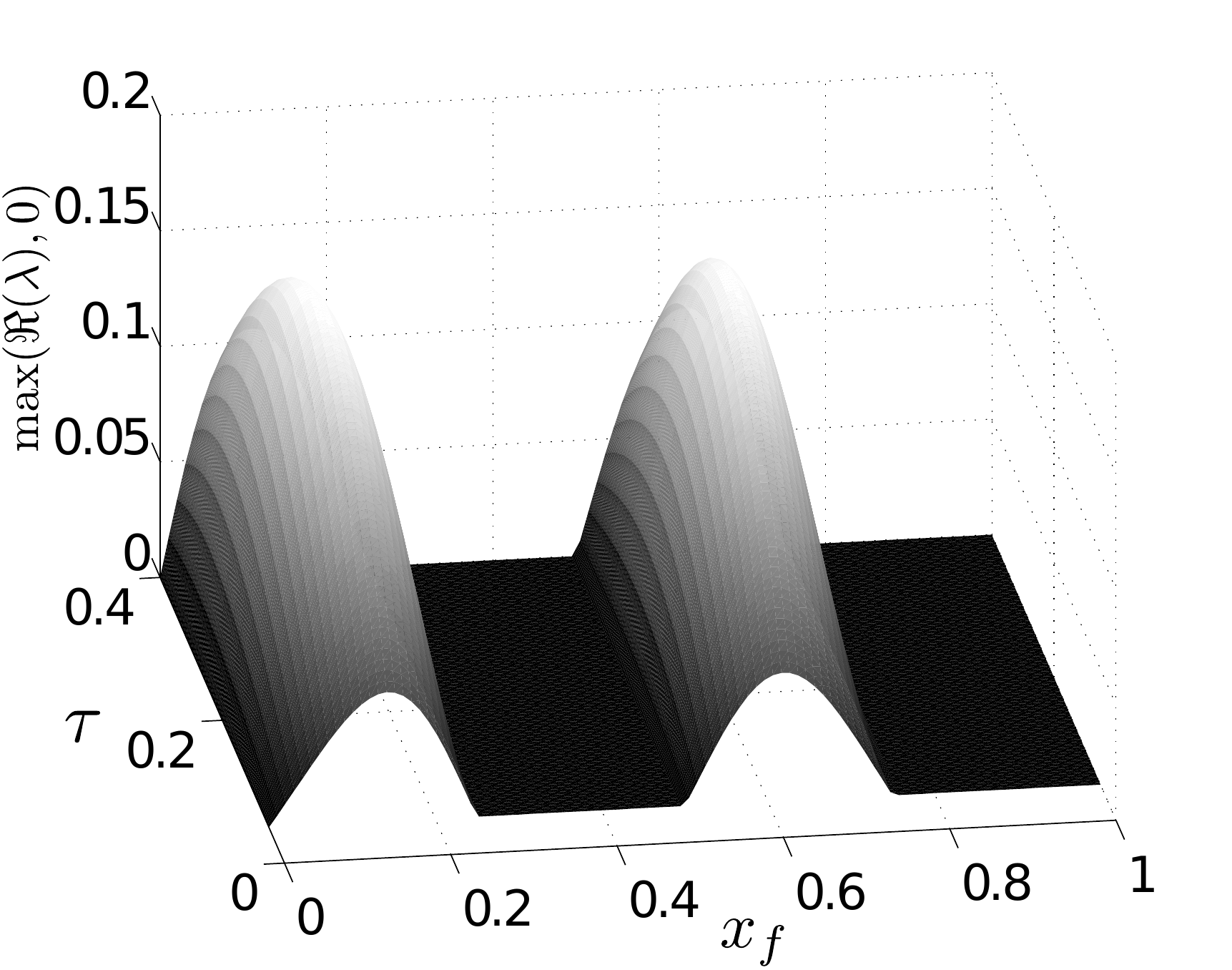}}
  \subfloat[$\hbox{imag}(\lambda) \approx 3\pi$]{\includegraphics[width = 0.33\textwidth] {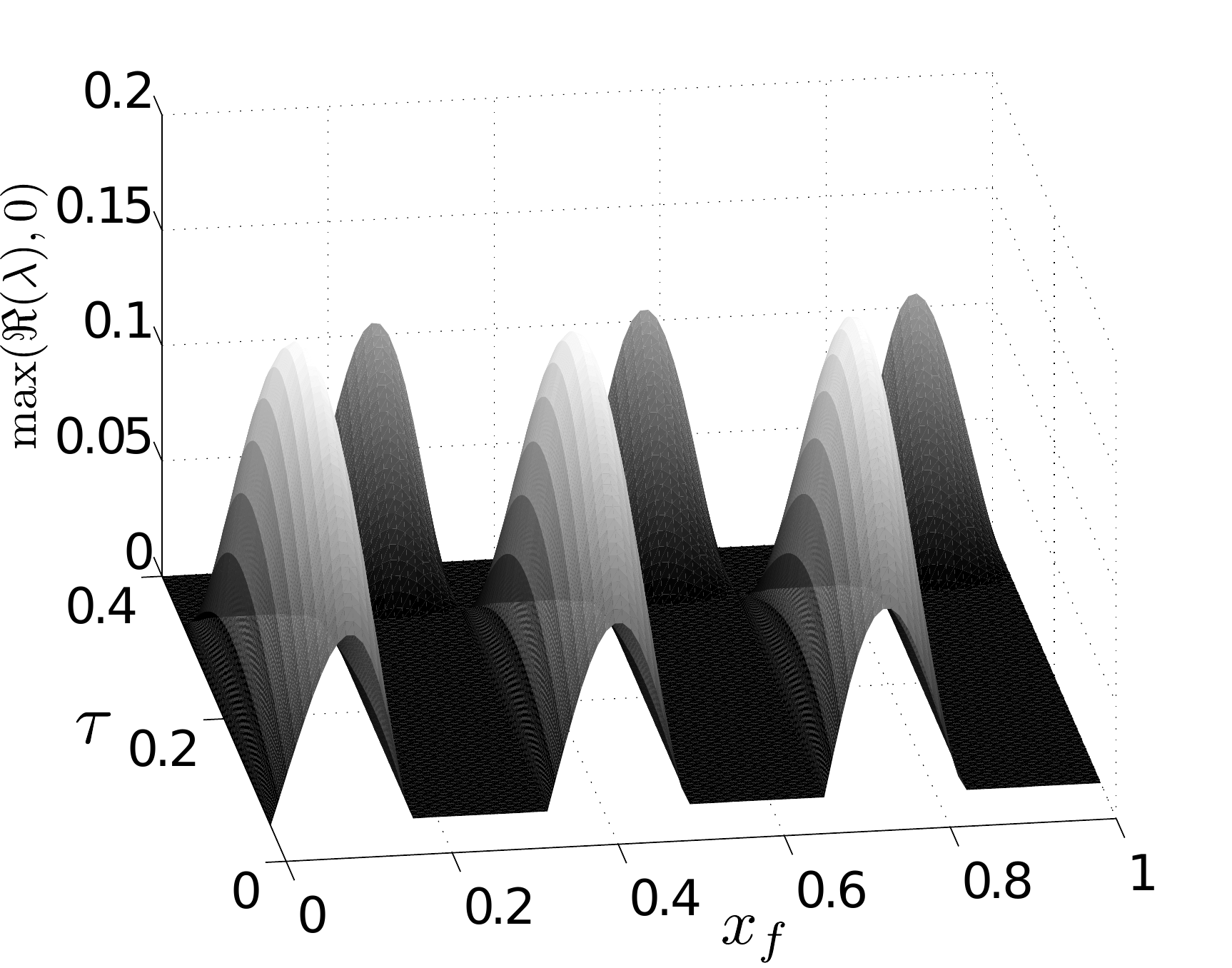}}
  \caption{The region of instability for the first three frequency modes ($K = 0.8$) for the analytical solution.}
  \label{fig:SR}
  \centering
{\includegraphics[width = 0.9\textwidth] {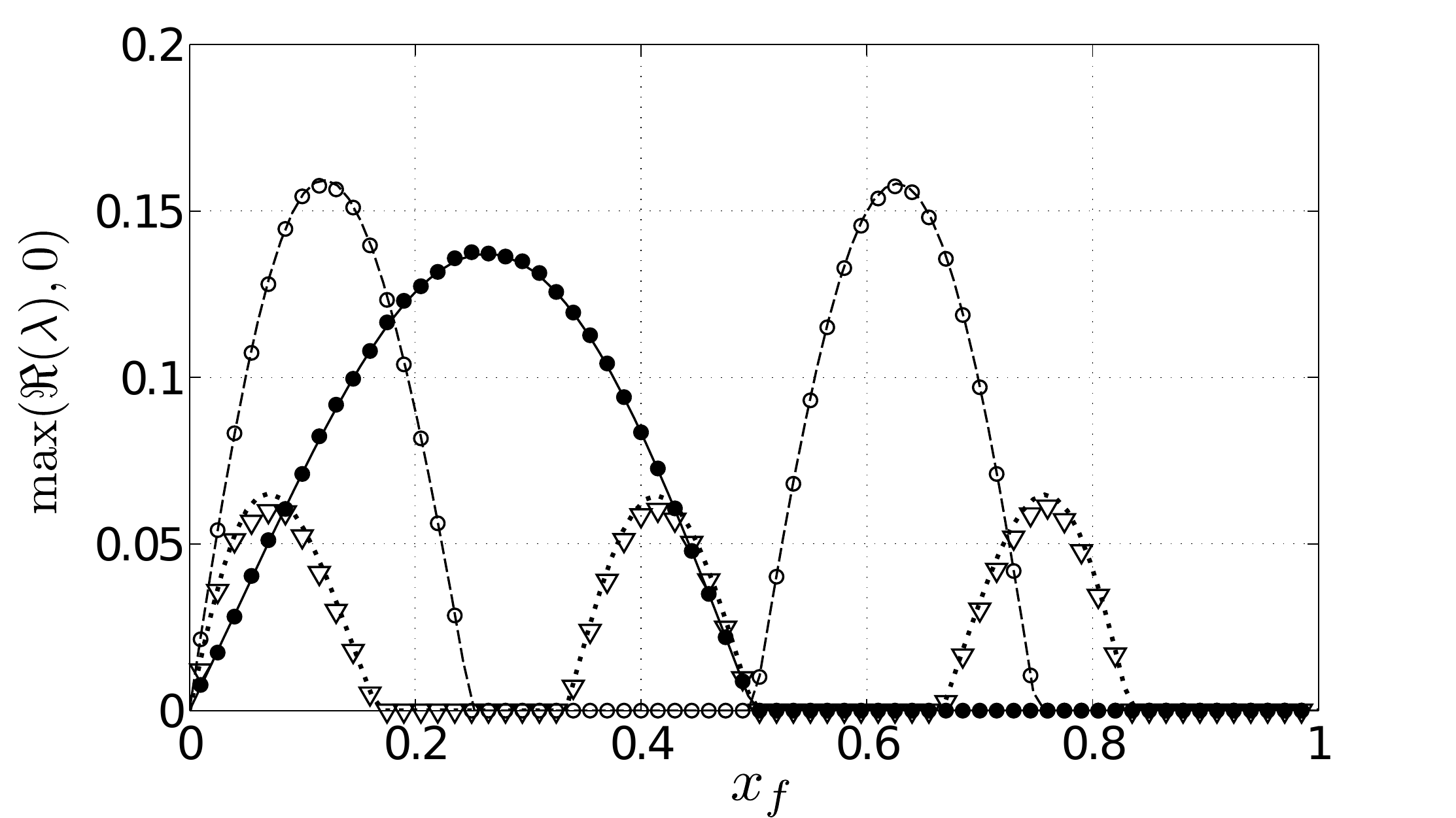}}
  \caption{Growth-rates of the first \red{three} frequencies; comparison of the high-fidelity and the analytical solution, $\tau = 0.3$ \& $K = 0.8$. $\hbox{imag}(\lambda) \approx \pi$: -----, analytical; $\bullet$, High-fidelity. $\hbox{imag}(\lambda) \approx 2\pi$: $---$, analytical; $\circ$, high-fidelity; $\hbox{imag}(\lambda) \approx 3\pi$: $\cdots$, analytical; $\Delta$, High-fidelity.}
 \label{fig:SR2}
\end{figure}

Figure~\ref{fig:ES} also shows that the high-fidelity system produces the best agreement with the analytical spectrum. The predictions of the spectra can be better illustrated by comparing the parametric stability diagrams of the high-fidelity approach to the analytical solution. The parametric instability diagram of the analytical solution is shown in Figure~\ref{fig:SR}, where the real part of the eigenvalue, i.e. the growth-rate, corresponding to the three low-frequency modes of $\hbox{imag}(\lambda) = \pi$, $2\pi$ and $3\pi$ is plotted in the parametric space of $\left ( x_f,\tau \right )$. The strength of the heater is kept constant at a value of $K = 0.8$. This figure demonstrates that depending on the location of the heater and the value of the delay all or none of these three low-frequency modes can be unstable. The predictions of the high-fidelity solution in the limit of low damping and at a fixed plane of $\tau = 0.3$ is compared to the analytical results in figure~\ref{fig:SR2}. The growth-rates from the high-fidelity approach agree well with the results of the analytical parametric stability diagram. However, the diagrams which are relevant to the experimental conditions are obtained for higher damping coefficients, where a traveling wave solution does not exist and no analytical solution is therefore available. The effect of damping on the stability of the system will be demonstrated in the following section. 

\subsection{Exponential growth of the time-integrated signal}
\label{ss:verification}
In the previous subsections, we established that the high-fidelity solution produced the closest agreement with the analytical solution in the limit of no damping and suffered the least from numerical errors. At this stage, we will first derive the instability diagrams of the linearized system with the damping coefficients described in \S~\ref{sec:GE}, which would reproduce experimental conditions. In addition, predictions of the linear stability analysis are verified by nonlinear calculations in the low-amplitude regime, corresponding to the initial growth of the signal.

\begin{figure}
  \centering
  \subfloat[Instability diagram. (red), $\hbox{imag}(\lambda) \approx \pi$; (blue), $\hbox{imag}(\lambda) \approx 2\pi$.]{\includegraphics[width = 0.6\textwidth] {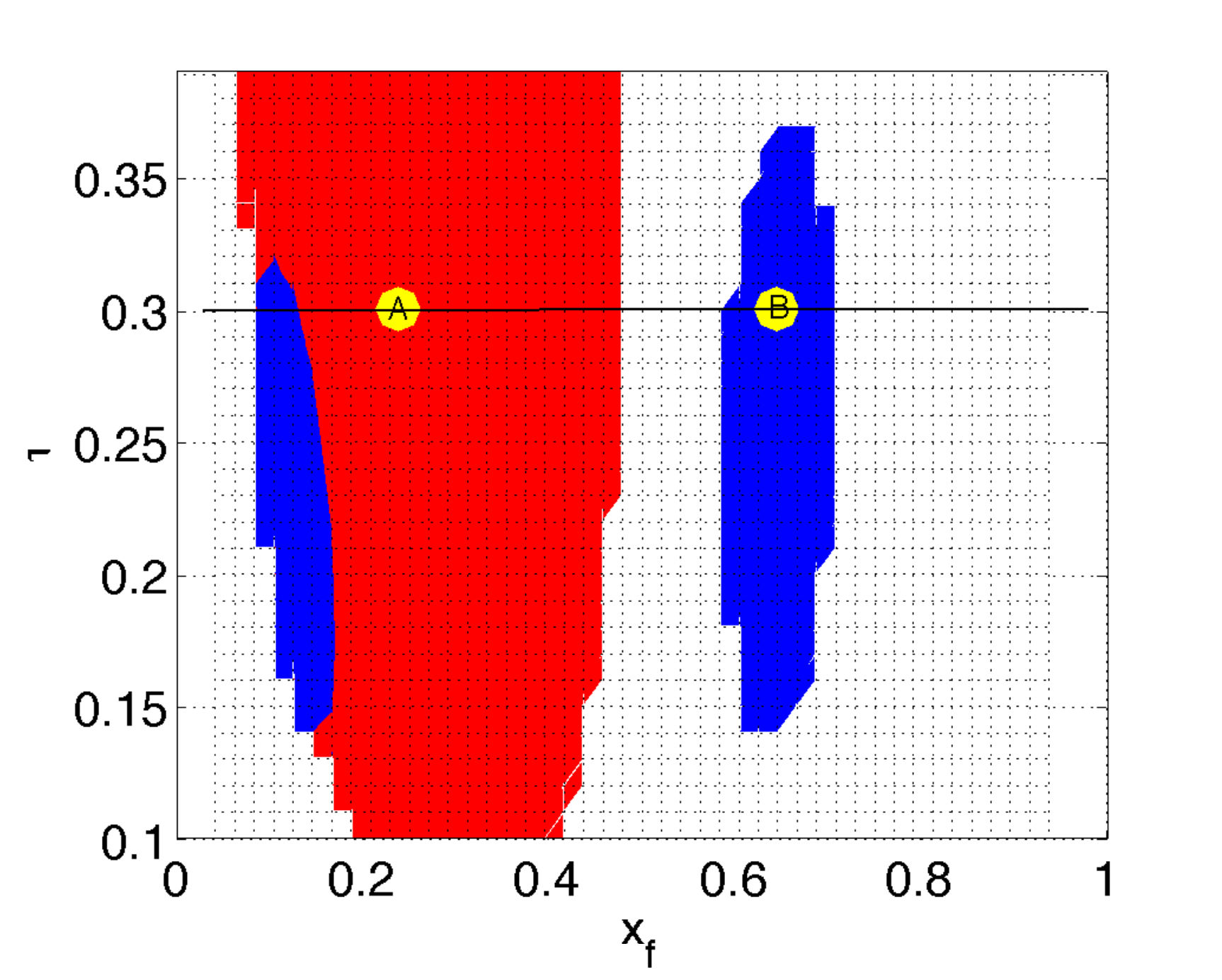}}\\
  \subfloat[Time-series, Case A]{\includegraphics[width = 0.33\textwidth] {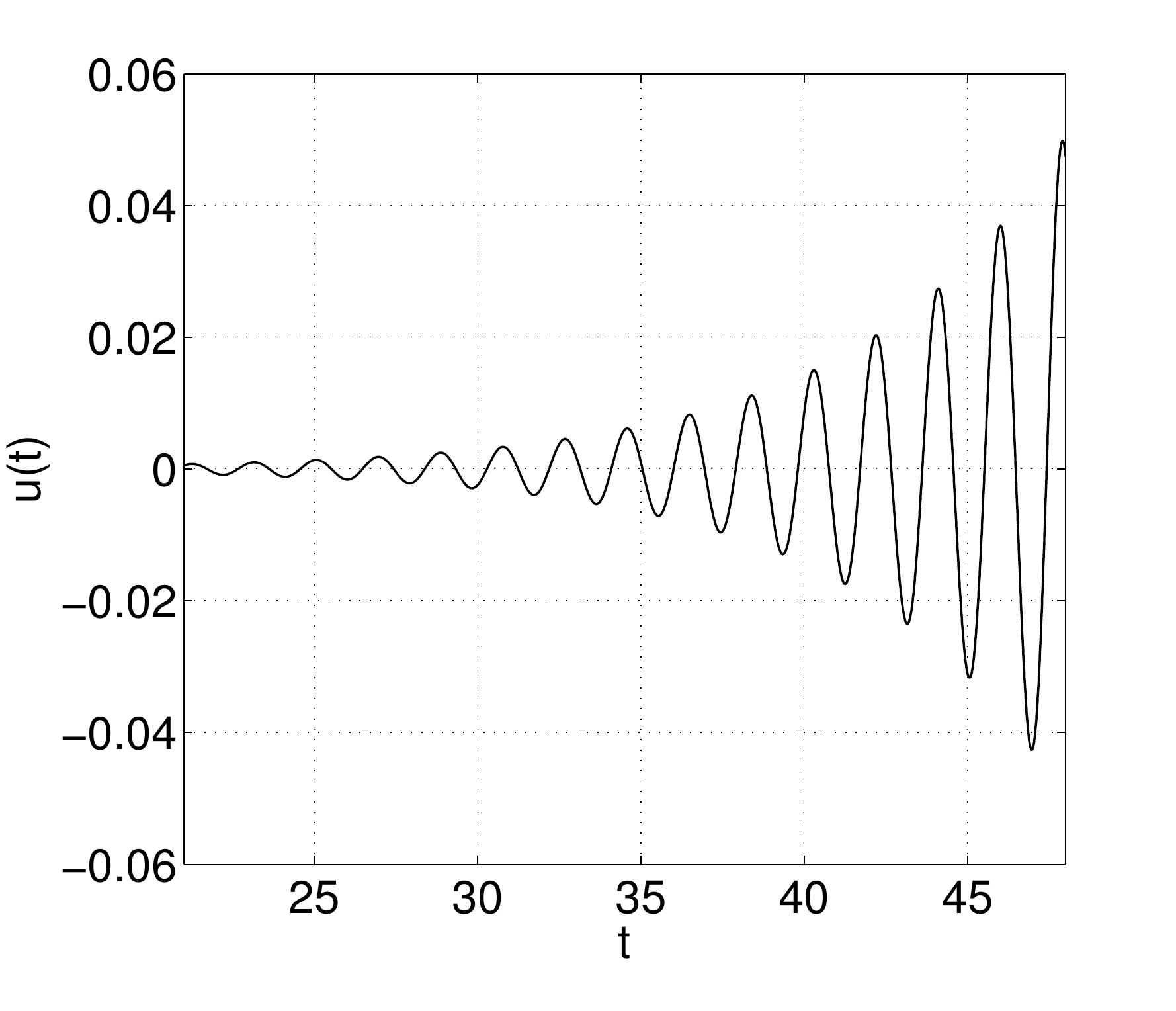}}
   \subfloat[Eigenvector, Case A]{\includegraphics[width = 0.33\textwidth] {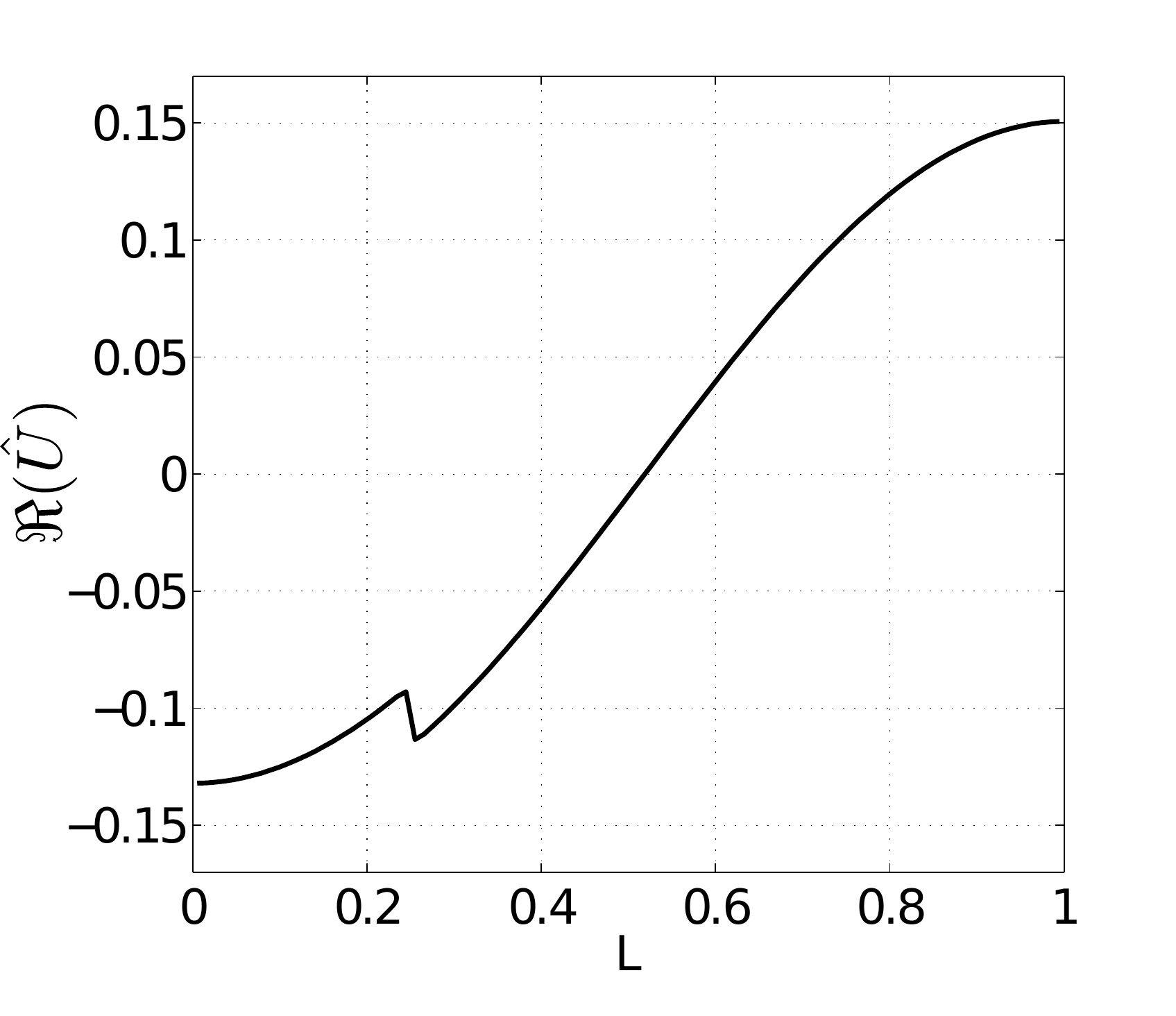}}
  \subfloat[Spatial distribution, Case A]{\includegraphics[width = 0.33\textwidth] {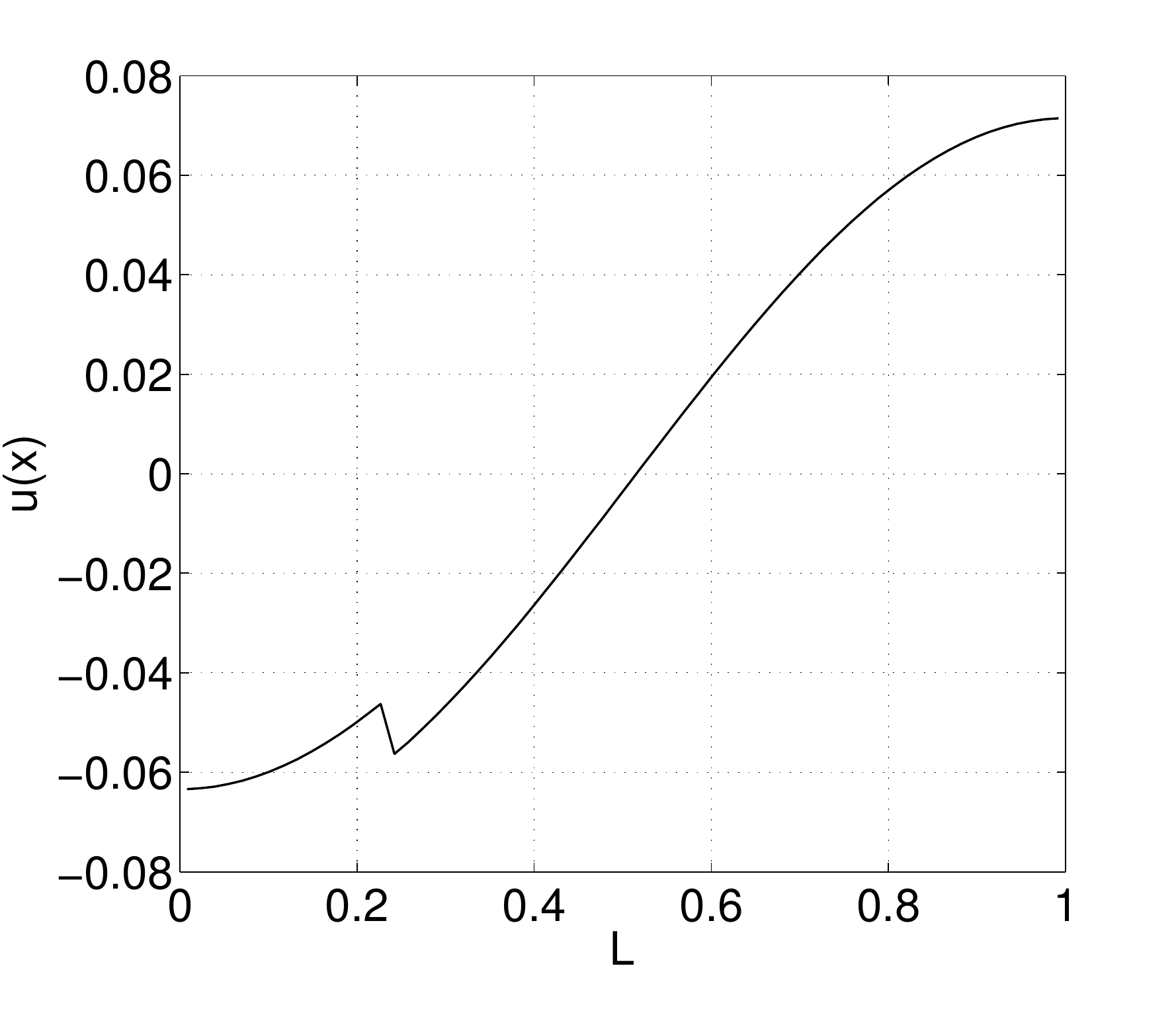}}\\
  \subfloat[Time-series, Case B]{\includegraphics[width = 0.33\textwidth] {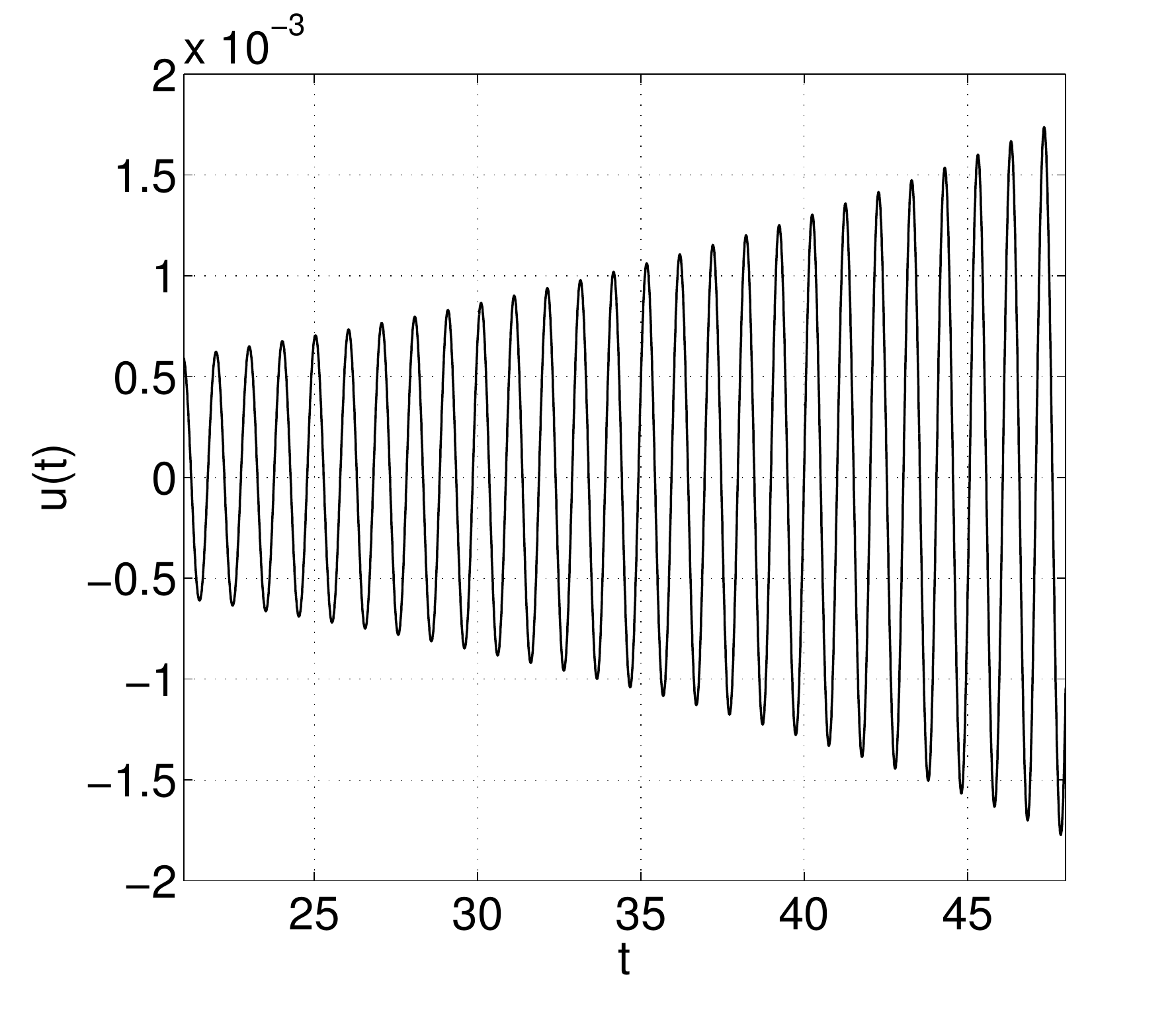}} 
  \subfloat[Eigenvector, Case B]{\includegraphics[width = 0.33\textwidth]{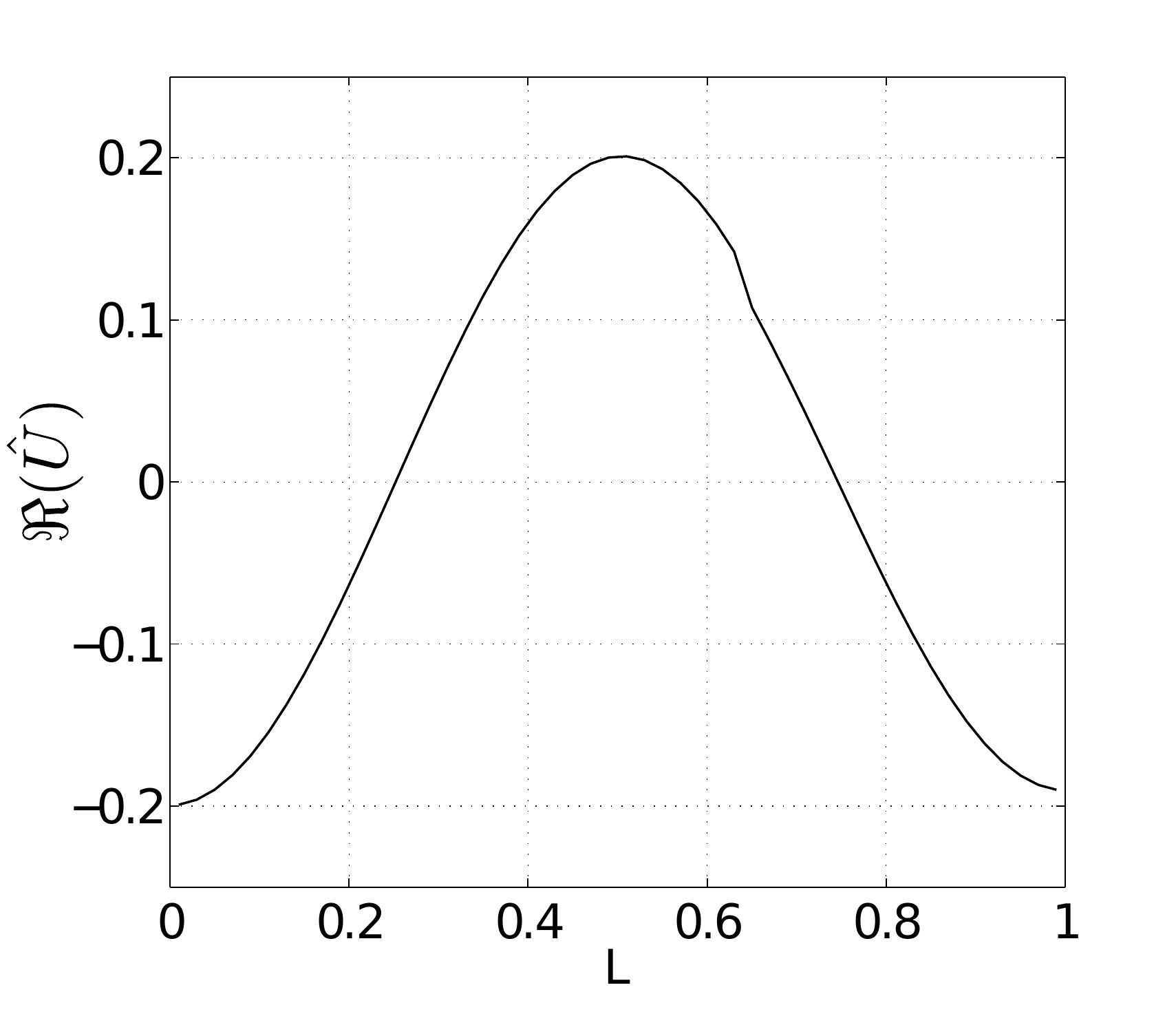}}
  \subfloat[Spatial distribution, Case B]{\includegraphics[width = 0.33\textwidth]{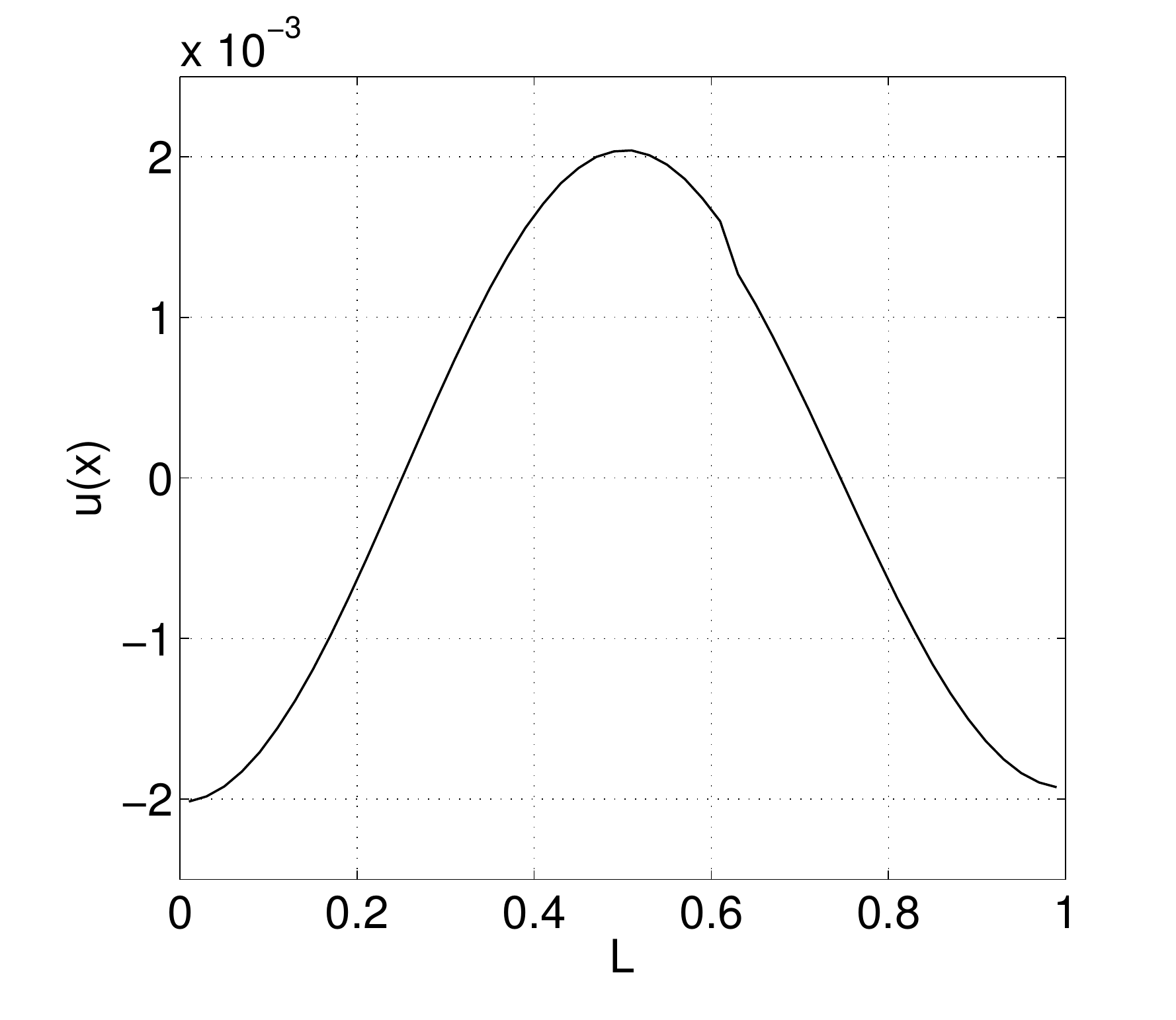}}
  \caption{Time signal from a probe located at $x/L = 0.2$, the eigenvector and the spatial distribution of the velocity signal in the linear regime. Case A: $\tau = 0.3$, $x_f = 0.25$, Case B: $\tau = 0.3$, $x_f = 0.64$.}
  \label{fig:vel_linear}
\end{figure}

The resulting parametric stability diagram for $K = 1.4$ is given in figure~\ref{fig:vel_linear}. This figure shows that, depending on the location of the heater or the value of the delay, only the first or the second mode is linearly unstable. This is in contrast to the instability diagram of figure~\ref{fig:SR}, where higher frequencies were unstable as well. This difference is due to the form of the damping which grows quadratically with the frequency and has a larger effect on the higher frequency modes. The prediction from the stability diagram of figure~\ref{fig:vel_linear}(a) also agrees with the experimental observations of \citet{matveev03}, where it was reported that in the first half of the tube the first frequency mode and in the second half the second frequency mode are most unstable. If the temporal evolution of the signal is dominated by the exponential growth of the linearly unstable mode, then the predictions from the instability diagram will determine the shape and frequency of the signal as well as the rate of the initial growth.

In order to compare the results of the instability diagram to the nonlinearly generated time series, two cases have been chosen; case A with parameters, $x_f = 0.25$ and $\tau = 0.3$ such that the lowest frequency mode of $\hbox{imag}(\lambda) \approx \pi$ is linearly unstable, and case B, which corresponds to the parameter values $x_f = 0.64$ and $\tau = 0.3$, where the most unstable frequency is about $2\pi$. For both cases the strength of the heater is set at $K= 1.4$; the higher strength in the heater allows for the discontinuity in the eigenfunctions to be clearly identifiable.

\begin{figure}
  \centering
  {\includegraphics[width = 0.6\textwidth] {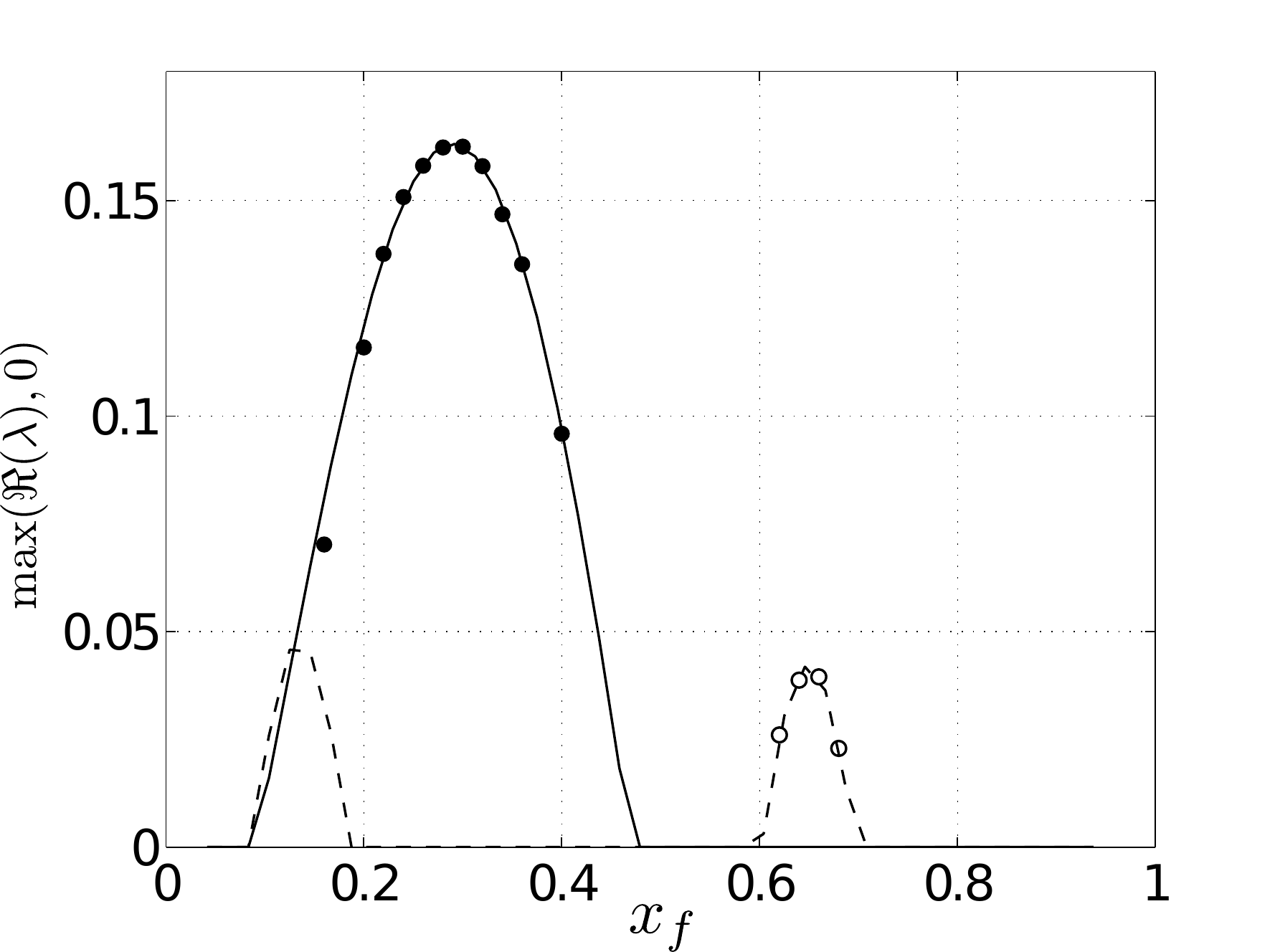}}
  \caption{Comparison of the growth-rates from the linear instability analysis and the time-integrated signal, $\tau = 0.3$ \& $K = 1.4$. $\hbox{imag}(\lambda) \approx \pi$: -----, linear; $\bullet$, nonlinear. $\hbox{imag}(\lambda) \approx 2\pi$: $---$, linear; $\circ$, nonlinear.}
  \label{fig:GR}
\end{figure}

In the early stages, the result of the fully nonlinear high-fidelity system is compared to the predictions from the instability diagram. Note that the instability diagram has been obtained for a linearized source term given by (\ref{eq:LST}). Therefore, for the assumptions of linearity to be valid, the evolution of the velocity signal from the initial stage, corresponding to the region of exponential growth, is considered. The time-series at a single probe location is shown in figures~\ref{fig:vel_linear}(b) and (e).  Figures~\ref{fig:vel_linear}(d) and (g) show the spatial shape of the velocity signal at a single point in time. The spatial signal coincides with the shape of the eigenvector of the first and second modes extracted from the linear stability analysis, presented in figure~\ref{fig:vel_linear}(c) and (f), suggesting that the most unstable wave detected through the linearized system governs the shape of the time-series, as expected. The location of the heater can be detected by the appearance of the discontinuity in the spatial mode. In order to verify whether, in the small-amplitude regime, the pressure and the velocity signal of the acoustic waves grow as predicted by linear analysis, the linear growth-rates are compared to the growth-rates that are measured through the time-integrated signal; the results are shown in figure~\ref{fig:GR}. This figure shows that in the first half of the duct the first mode $\hbox{imag}(\lambda) \approx \pi$ is the most unstable frequency which dominates the exponential growth of the signal in the nonlinear calculation. In contrast, in the second half of the duct the second frequency $\hbox{imag}(\lambda) \approx 2\pi$ is the most unstable mode and the growth-rate of this mode from the time-integrated calculation matches the linear prediction, suggesting that in both cases the signal grows exponentially as suggested by the linear stability diagram. 

\section{The dynamics governing the limit cycle}
\label{sec:fullanalysis}
In this section, the performance of each method is assessed by evaluating the time-series as they traverse from the linear to the fully nonlinear regime. In particular, the effect of the strength of the heater on the shape of the limit cycle is examined, illustrating the importance of coupling as the strength of the heater is increased. In the linear stage, the unstable frequencies and their respective spatial shapes are extracted by the linear stability method of \citet{jarlebring08} applied to the semi-discretized system of equations as described in \S~\ref{sec:CDM}. Further development of the signal into the nonlinear regime modifies the mean and the unstable frequencies that dominate the limit cycle. Coupling between different frequency modes could also cause higher harmonics of a specific frequency to become significant. To verify whether this is the case for the thermoacoustic signal, we compare the limit cycles of the coupled Galerkin approach to those of the decoupled method. Moreover, comparison of the limit cycles produced by the high-fidelity approach to the coupled Galerkin solutions highlights the importance of correctly accounting for the discontinuity.  Finally, we extract dominant frequencies from an established limit cycle and compare the results to the frequencies and modal shapes of the linear regime. 

\begin{figure}
  \centering
  {\includegraphics[width = 0.6\textwidth] {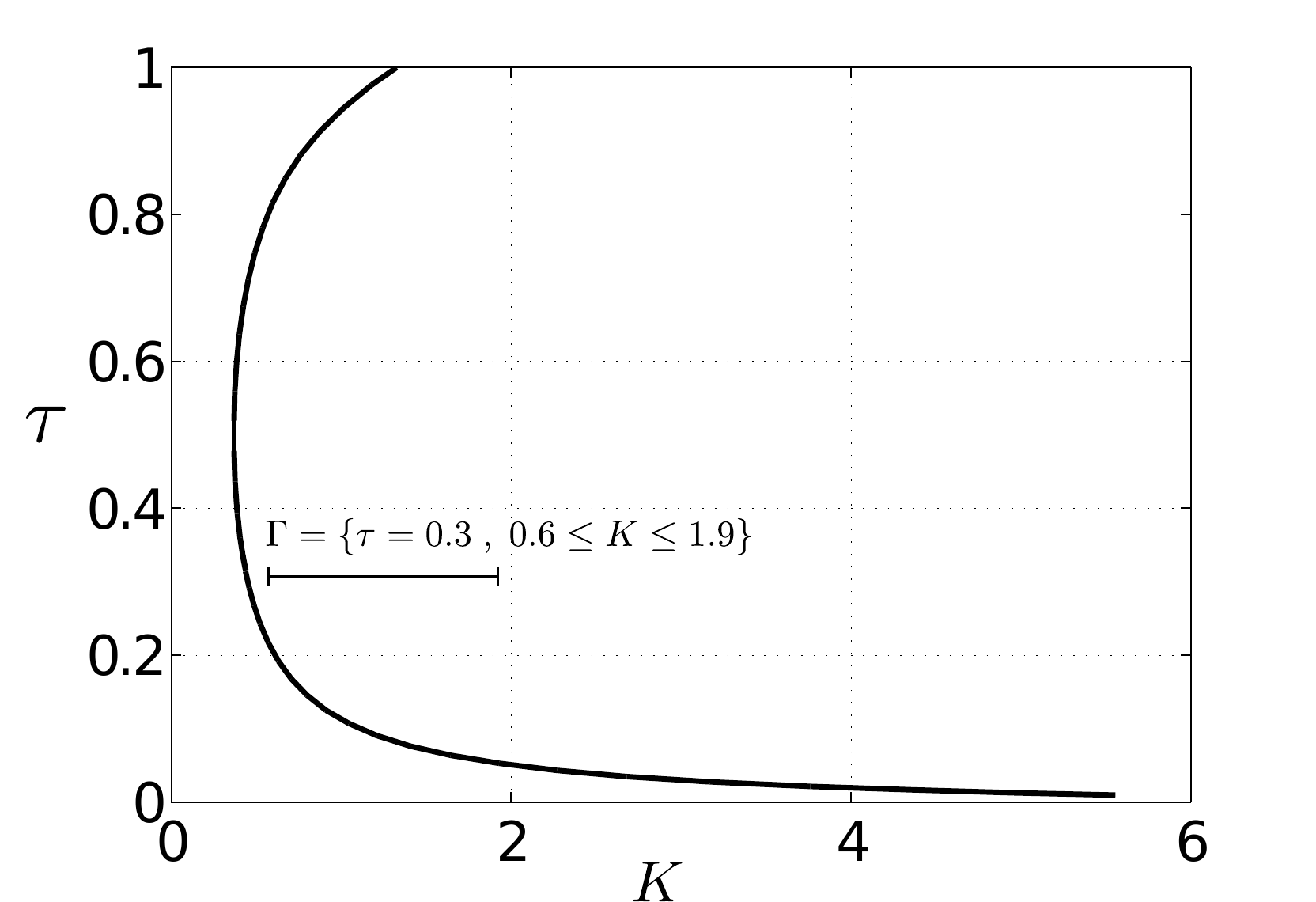}}
  \caption{Neutral curve of the first frequency mode $\hbox{imag}(\lambda) \approx  \pi$ at $x_f = 0.25$.}
  \label{fig:NC}
\end{figure}

We wish to determine whether the most unstable frequency carries through to the nonlinear regime and whether it remains dominant in the resulting limit cycle. To this end, the location of the heater is fixed at $x_f = L/4$, while the remaining system parameters, $[\tau\; , \; K]$, are allowed to change. The neutral curve of the first acoustic modes, corresponding to the frequency $\hbox{imag}(\lambda) \approx \pi$, is plotted in figure~\ref{fig:NC}. This figure shows that for heater strengths of up to $K = 6$, the first acoustic mode is the only unstable mode of the system. In order to investigate the nonlinear growth of the signal and the effect of the heat release rate on the resulting limit cycle, the strength of the heater is changed from $K = 0.6$, close to the boundary of linear instability, to $K = 1.9$ at a constant delay, resulting in the parameter space of $\Gamma= \left[ \tau = 0.3\; , \;0.5 \le K \le 1.9 \right]$, shown on the neutral curve of figure~\ref{fig:NC}. The heater is placed at $x_f = L/4$ since, based on the instability diagrams of the previous section, this location is the most unstable location for the first frequency mode. 

\begin{figure}
  \subfloat[Decoupled Galerkin]{\includegraphics[width = 0.33\textwidth] {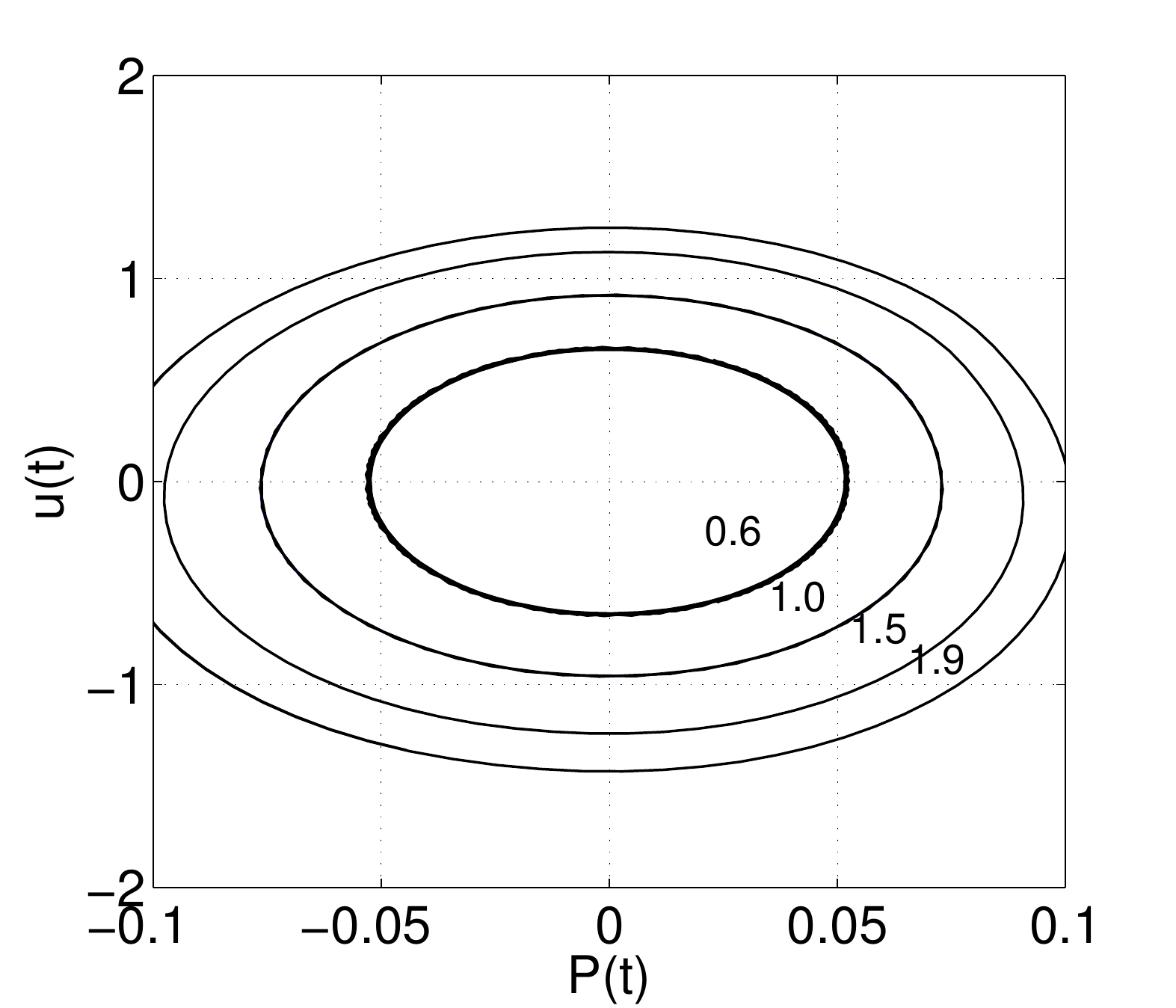}}
  \subfloat[Coupled Galerkin]{\includegraphics[width = 0.33\textwidth] {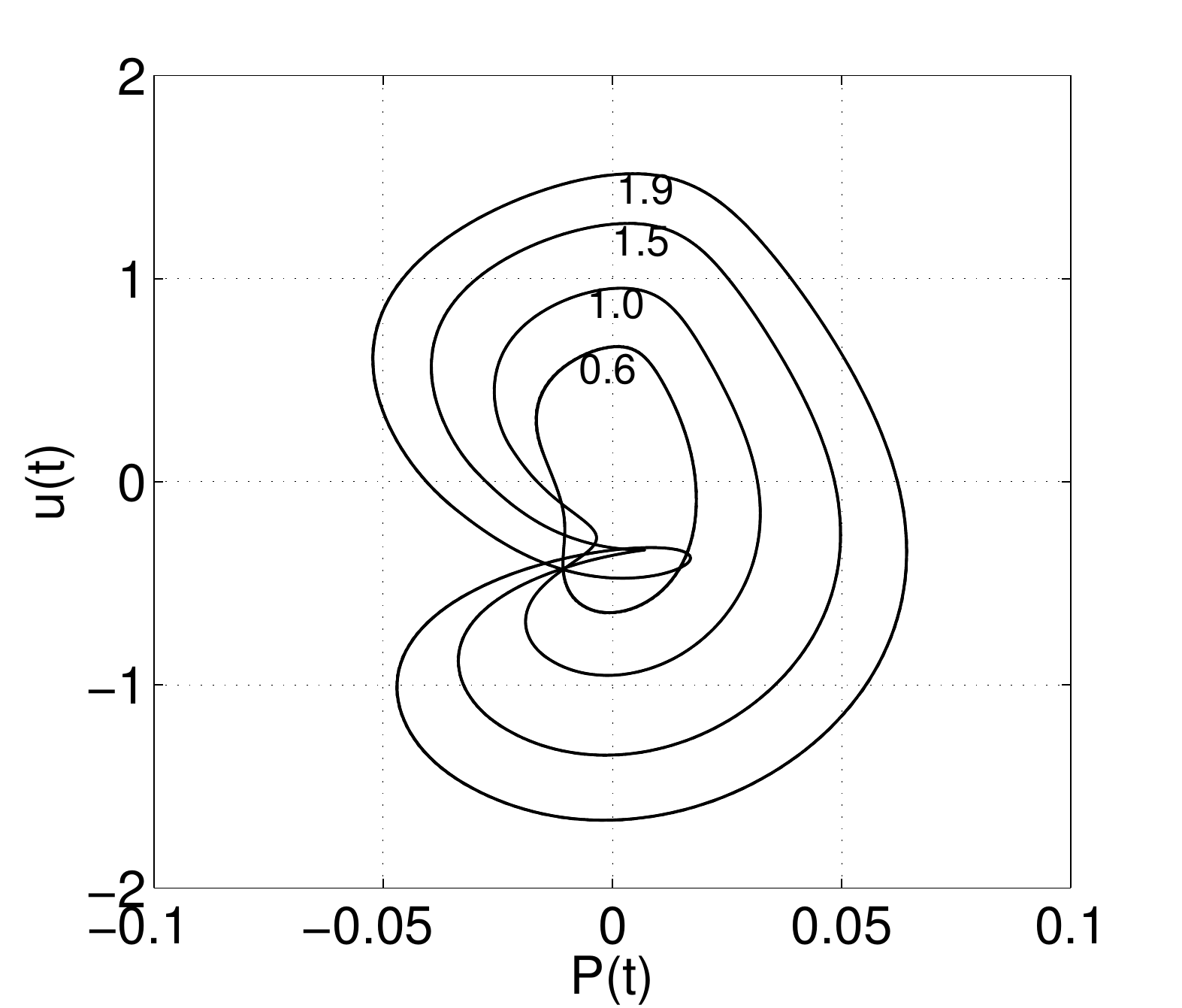}}
 \subfloat[High-fidelity]{\includegraphics[width = 0.33\textwidth] {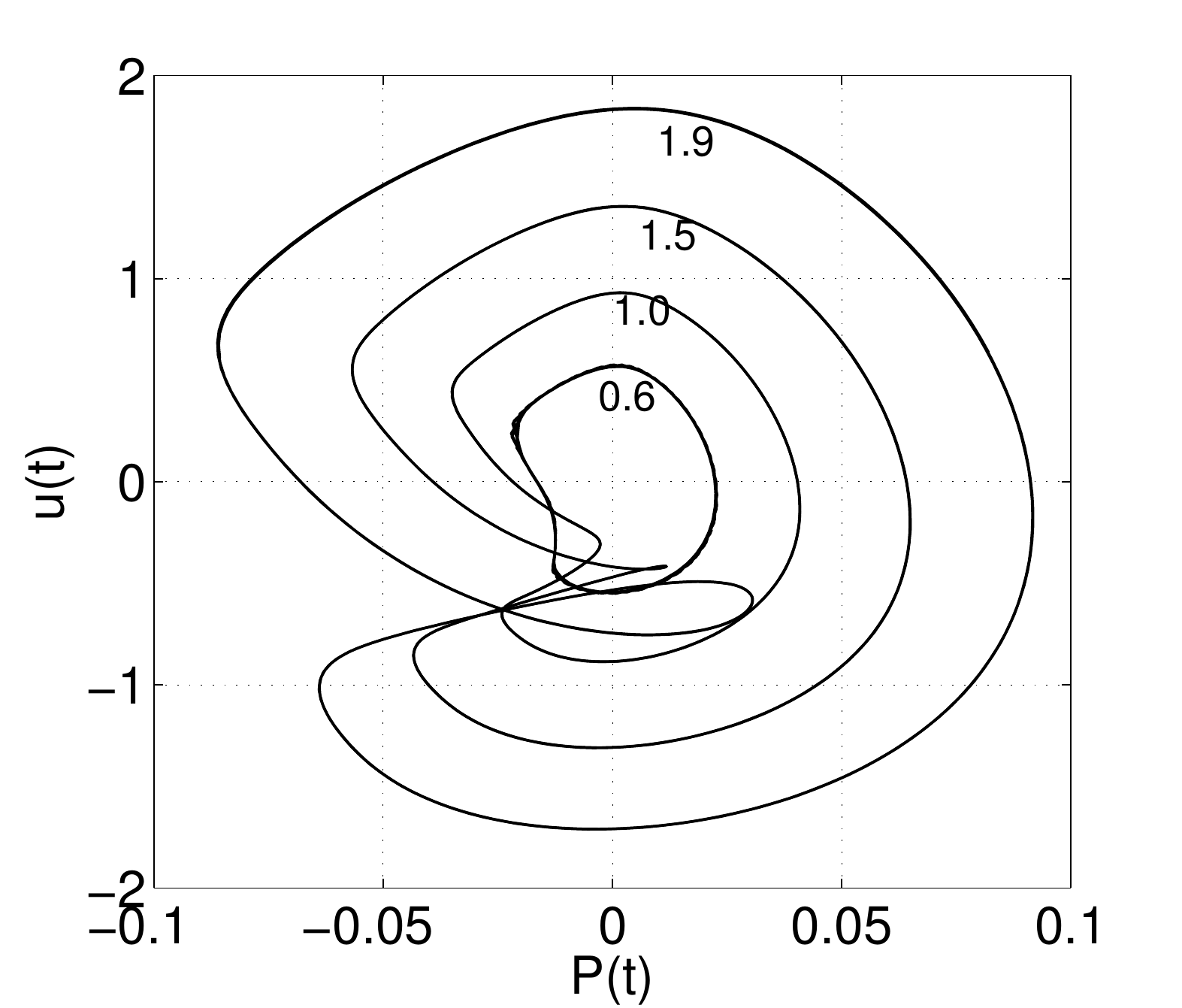}}
  \caption{The effect of the strength of the heater, $K$, on the final limit cycle. Comparison of the nonlinear high-fidelity solution and the decoupled Galerkin approach.}
  \label{fig:K_LC}
\end{figure}

The evolution of the limit cycle with varying heater strength is shown in figure~\ref{fig:K_LC} for the decoupled and coupled Galerkin scheme as well as the high-fidelity solutions. For lower values of $K$, all methods produce a unimodal limit cycle. However, the amplitude and shape of the coupled Galerkin approach is closer to the high-fidelity solution. The decoupled Galerkin solution shows limit cycles with higher amplitudes for pressure in comparison to the other methods. This distinction becomes more clear as the strength of the heater is increased. The limit cycles developed through the high-fidelity and the coupled Galerkin solutions gradually evolve into a bimodal cycle, while the form of the limit cycle from the decoupled Galerkin solution remains the same, suggesting that increasing the strength of the heater affects the coupling of the modes causing the higher harmonic of the low-frequency mode to become unstable. Note that, as shown in the neutral curves of figure~\ref{fig:NC}, K varies in the range where only the first frequency mode is linearly unstable. As a result the appearance of the second frequency in the limit cycle is only due to the nonlinear effects during the evolution of the signal. The decoupled Galerkin scheme does not account for this effect, since the evolution of each frequency is considered independently from other frequencies. Although the coupled Galerkin approach accounts for this coupling, it underpredics the amplitude and ultimately the shape of the limit cycle, which is due to numerical errors introduced by high-frequency oscillations, as shown in \S~\ref{ss:SandM}.

\begin{figure}
 \centering
  \subfloat[Decoupled Galerkin, $K = 0.6$] {\includegraphics[width = 0.48\textwidth] {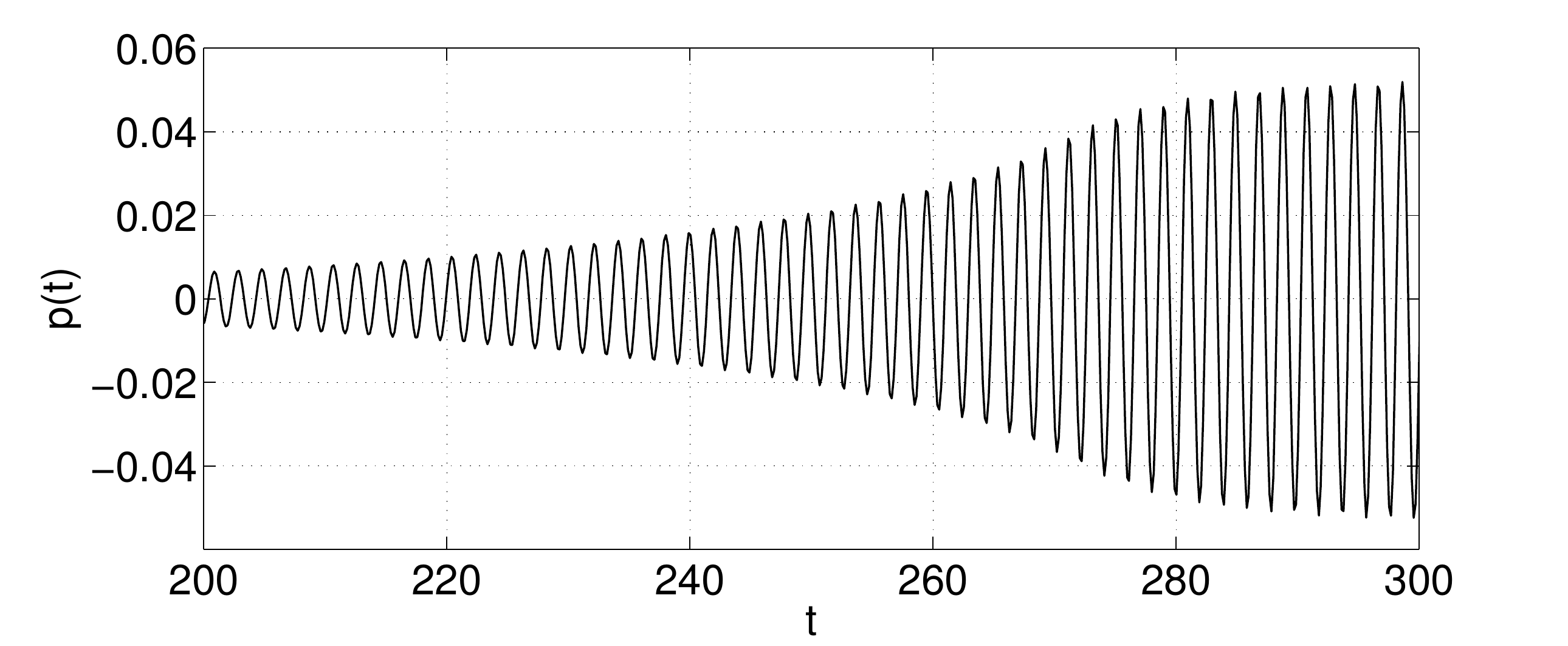}}
 \subfloat[Decoupled Galerkin, $K = 1.9$] {\includegraphics[width = 0.52\textwidth] {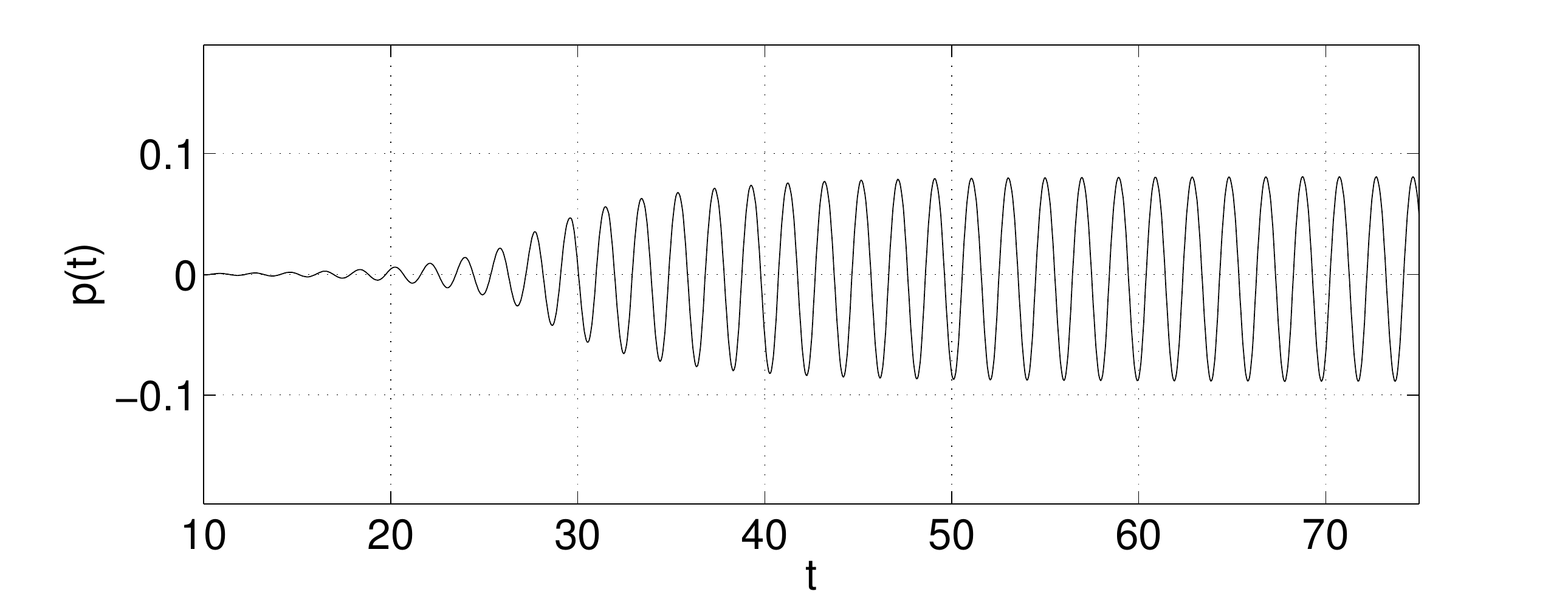}}\\
  \subfloat[High-fidelity, $K = 0.6$]{\includegraphics[width = 0.48\textwidth] {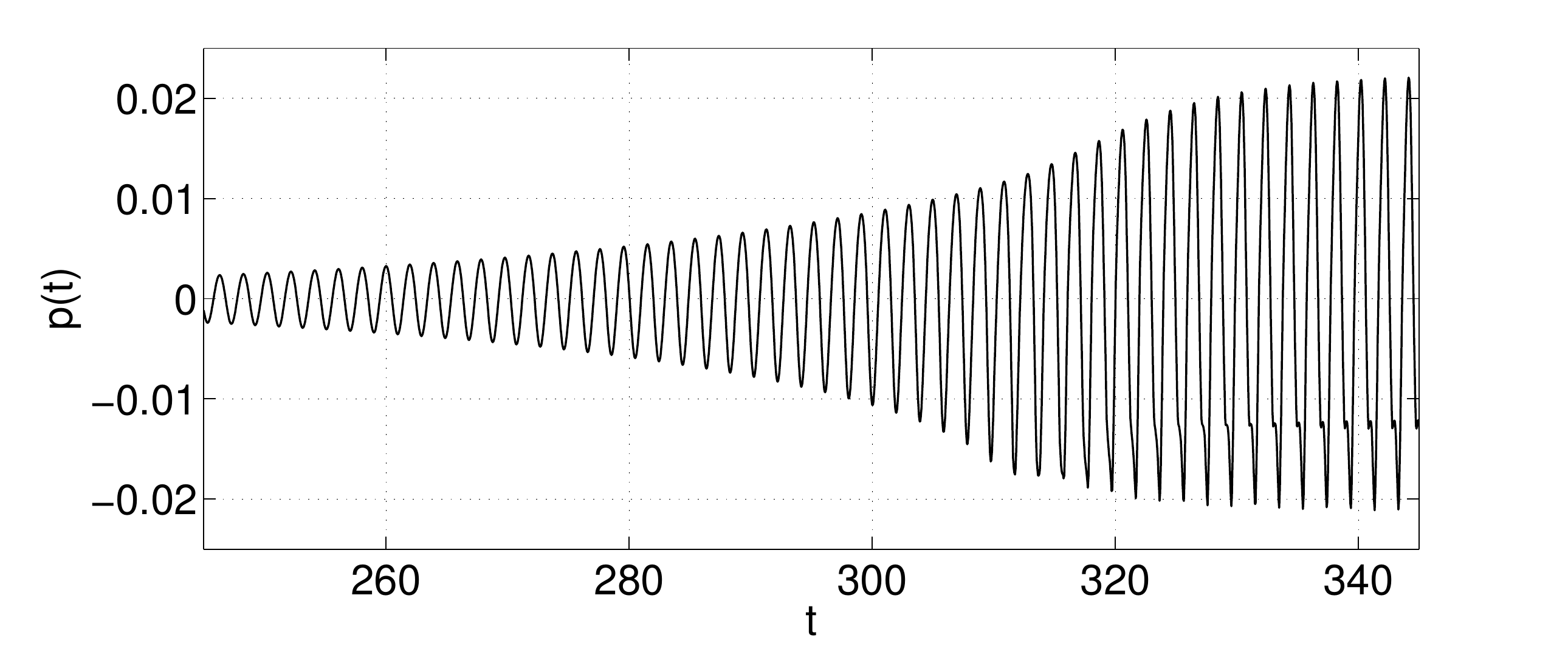}}
 \subfloat[High-fidelity, $K = 1.9$]{\includegraphics[width = 0.52\textwidth] {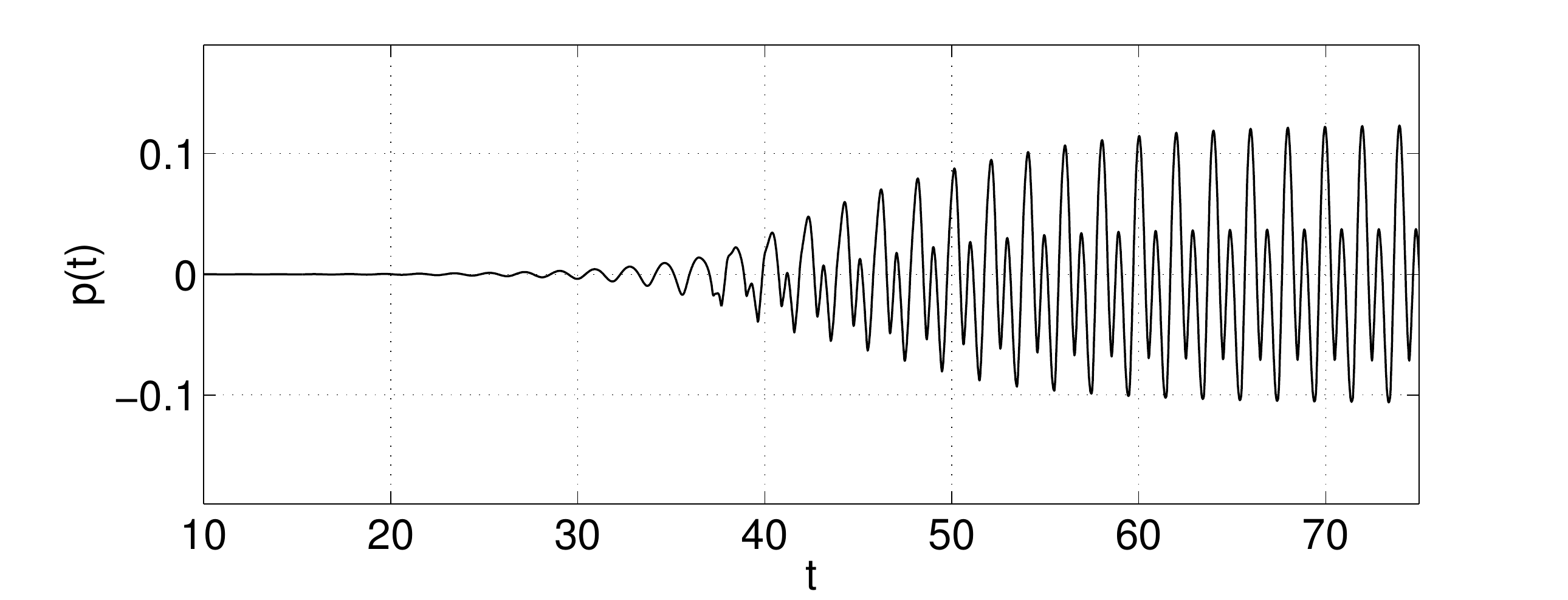}}
 \caption{Time-series of the pressure signal at $x = L/8$, comparison between the decoupled Galerkin and the high-fidelity solutions.}
 \label{fig:signal}
\end{figure}

The nonlinear evolution of the pressure signal for two heater strengths of $K = 0.6$ and $K = 1.9$ compared in figure~\ref{fig:signal}. This figure shows that for the lower value of heater strength the evolution of the signal from small-amplitudes to the limit cycle is similar between the decoupled Galerkin and the high-fidelity solution. However, for the higher heater strength, the results of the high-fidelity approach suggests that the frequency of the most unstable wave, and therefore the dominant frequency of oscillations, changes during the passage from the linear to the nonlinear regime. 

Through linear stability analysis we have access to the unstable eigenfunctions and frequencies; however, the dominant modes in the nonlinear regime need to be extracted from the computed time-signal. We employ the dynamic mode decomposition \citep{schmid2010} technique as one way to extract dynamically important modes from the final limit cycle. The DMD formalism has perviously been used by \citet{mariappan2011} to analyze experimental investigations of thermoacoustic instabilities. DMD is a spectral data-analysis technique, which decomposes data into spatial modes appearing with a specific (complex) frequency in the dataset. Since the objective here is to extract the dominant spatial modes and their respective frequencies in the nonlinear regime (for a comparison with the dominant linear modes), DMD appears to be a suitable tool. As an example, if $u(x,t)$ is a flow variable, where $x$ is the spatial coordinate and $t$ is time, $u$ can be represented by
\begin{equation}
  u(x,t) = \sum_{n=1}^N a_n\exp{(\lambda_n t)}\phi_n(x), \qquad
  \lambda_n \in \mathbb{C} \label{eq:decomp}
\end{equation}
where $\phi_n$ stands for the spatial modes, $a_n$ for their amplitudes and $\lambda_n$ for their frequencies. If the modes $\phi_n$ are normalized, the DMD-amplitudes, $a_n$, versus the extracted frequencies $\hbox{imag}(\lambda_n)$ can be used as a measure of dominance of each mode in the signal. Once the acoustic signal saturates and forms a limit cycle, few frequencies dominate the dynamics. This can be deduced by the form of the limit cycle itself, shown in figure \ref{fig:K_LC}. As a result, any feature extraction technique capable of extracting specific frequencies from the data, such as Fourier transforms, can be used to identify the dominant modes. We have chosen DMD in this study since it displays the least amount of spectral leakage \citep{sayadiCTR13}, and spatial modes with their respective frequencies, forming the modes of the nonlinear regime, can be extracted by analyzing a few data points of the saturated cycle.
\begin{figure}
  \centering
 {\includegraphics[width = 0.4\textwidth] {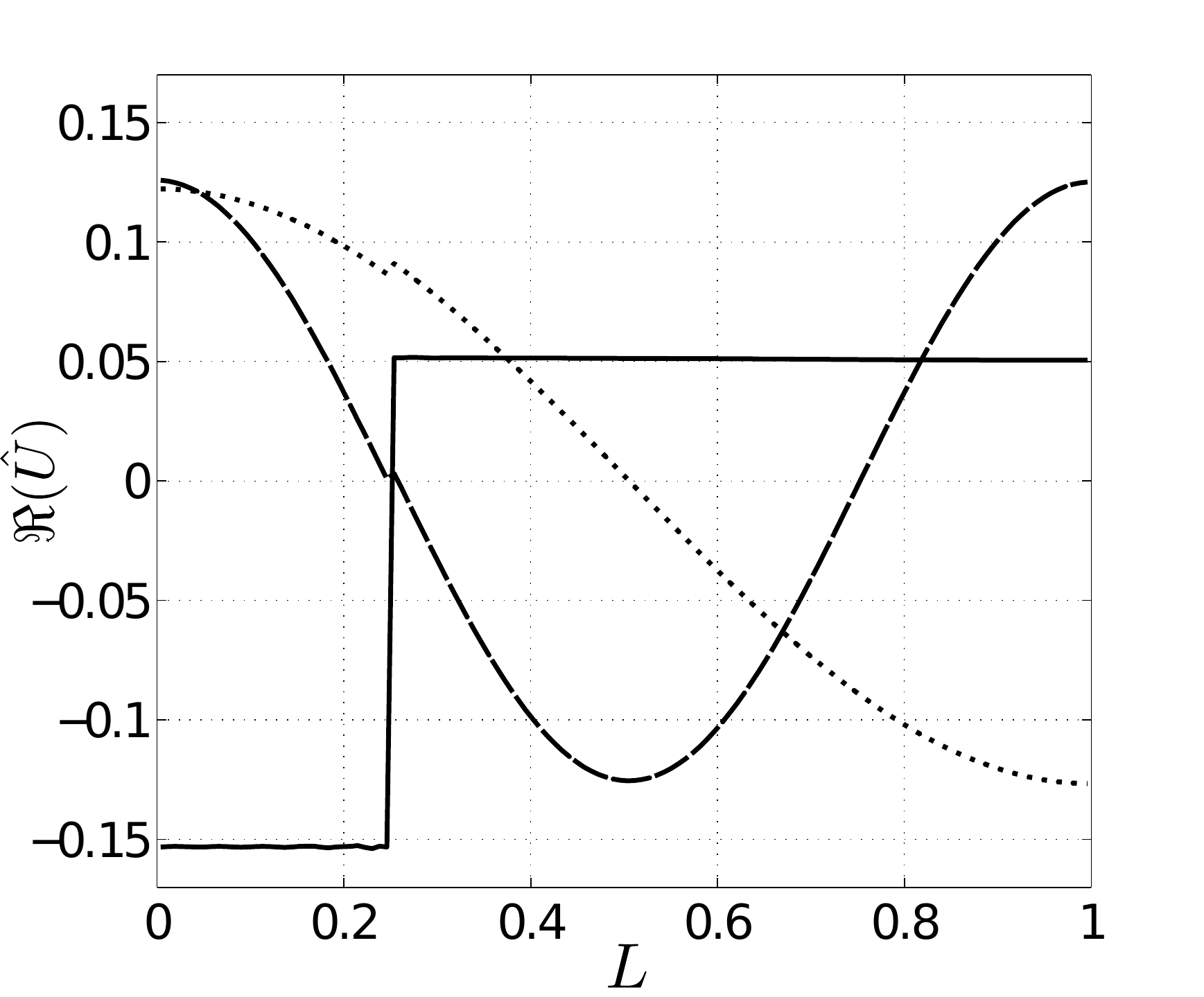}}
  \caption{The real part of the DMD eigenvectors. -----, $\hbox{imag}(\lambda) \approx 0$; $---$, $\hbox{imag}(\lambda) \approx \pi$; $\cdots$, $\hbox{imag}(\lambda) \approx 2\pi$.}
  \label{fig:DMD_mode}
  \subfloat[$K = 0.6$]{\includegraphics[width = 0.26\textwidth] {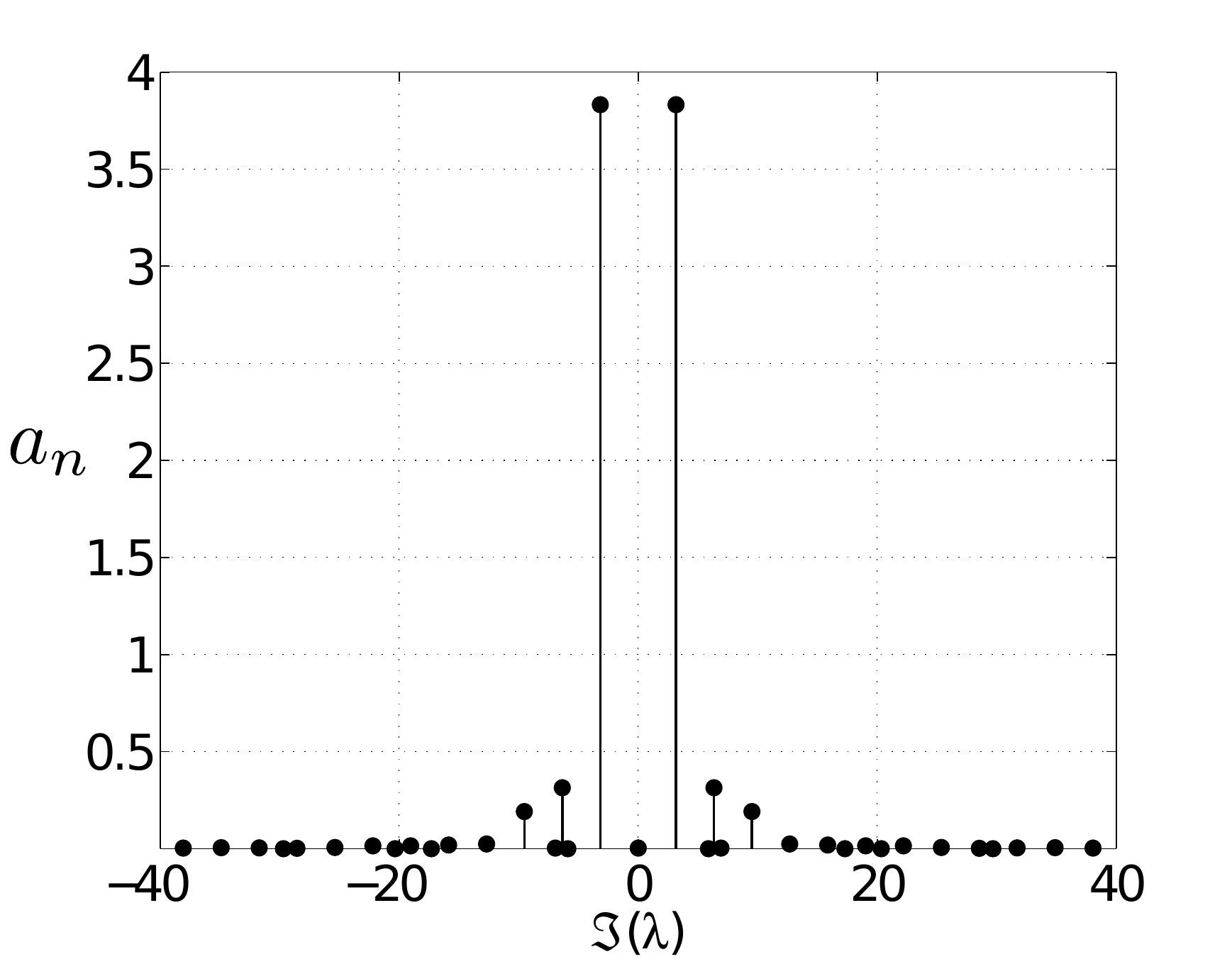}}
  \subfloat[$K = 1.0$]{\includegraphics[width = 0.26\textwidth] {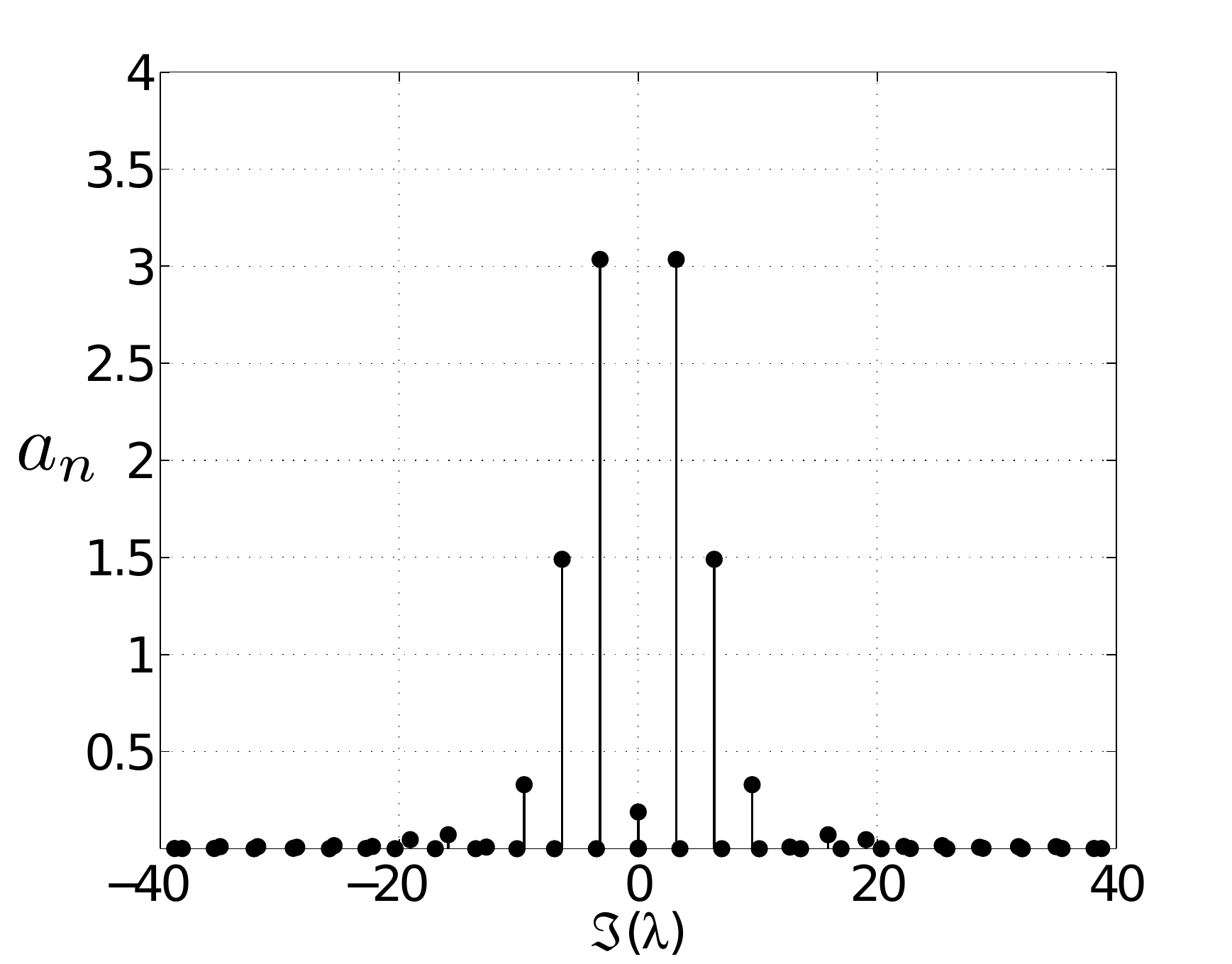}}
  \subfloat[$K = 1.5$]{\includegraphics[width = 0.26\textwidth] {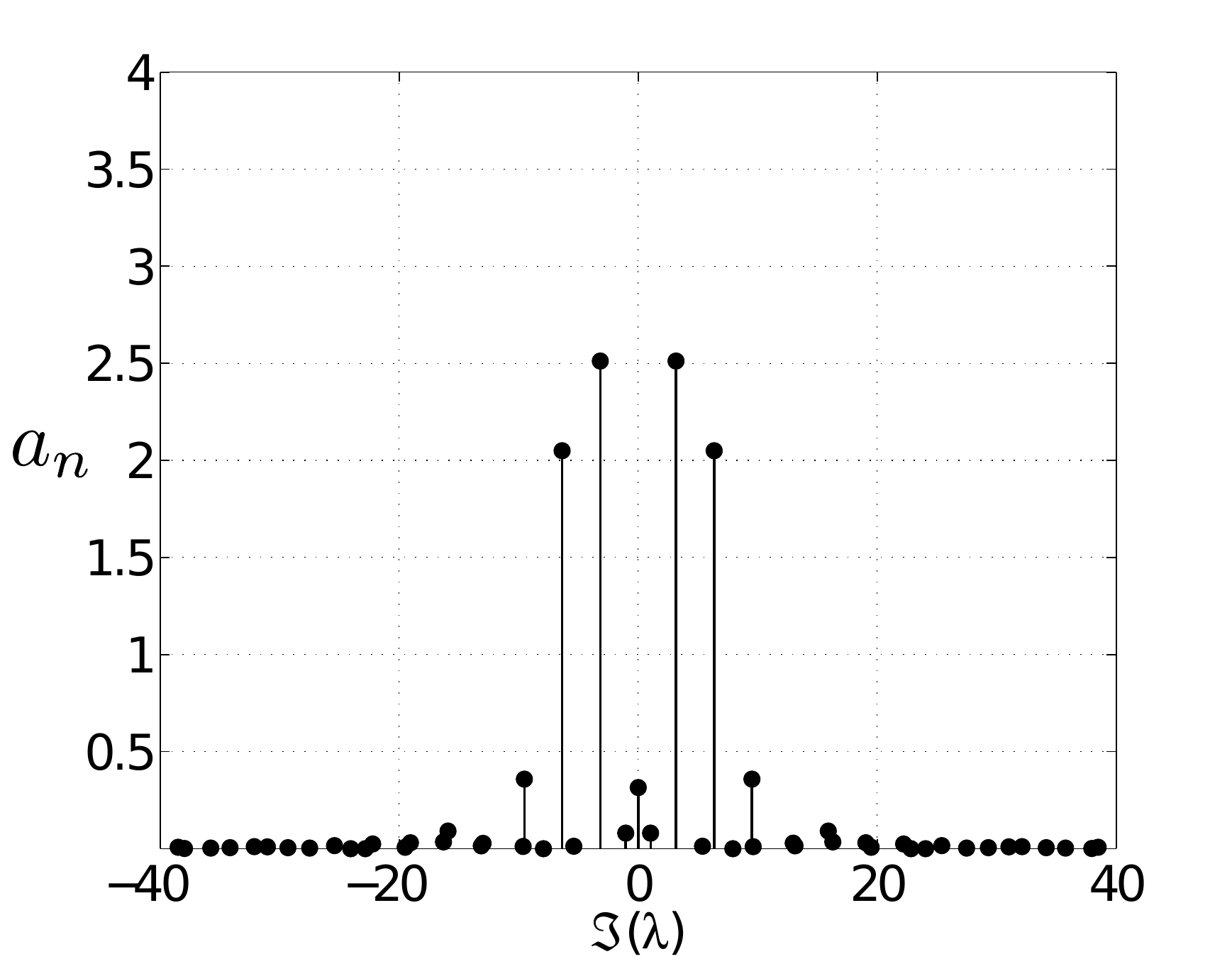}}
  \subfloat[$K = 1.9$]{\includegraphics[width = 0.26\textwidth] {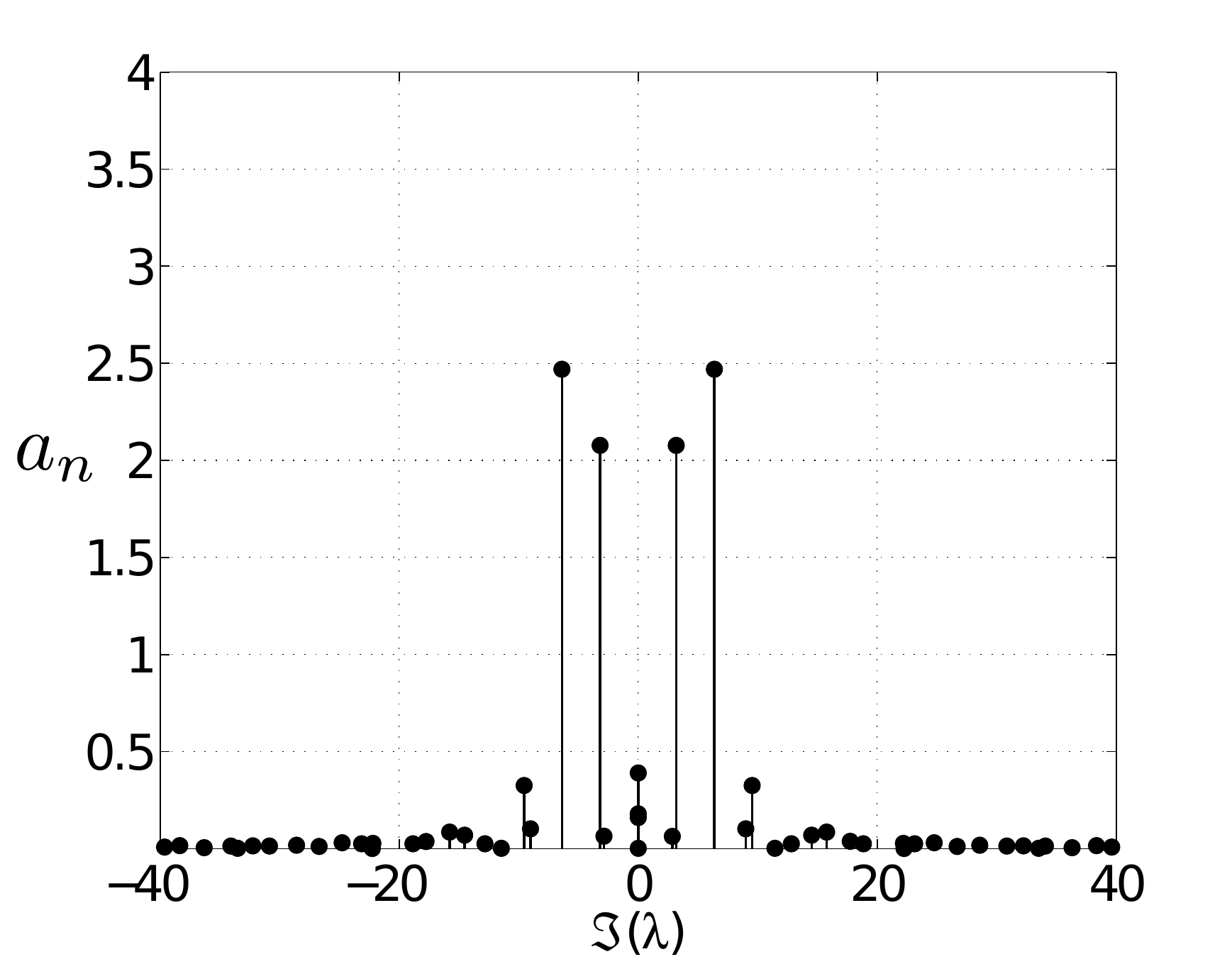}}
  \caption{DMD amplitude distribution, $\tau = 0.3$ and $x_f = 0.25$.}
  \label{fig:DMD_spec}
  \subfloat[$K = 0.6$]{\includegraphics[width = 0.26\textwidth] {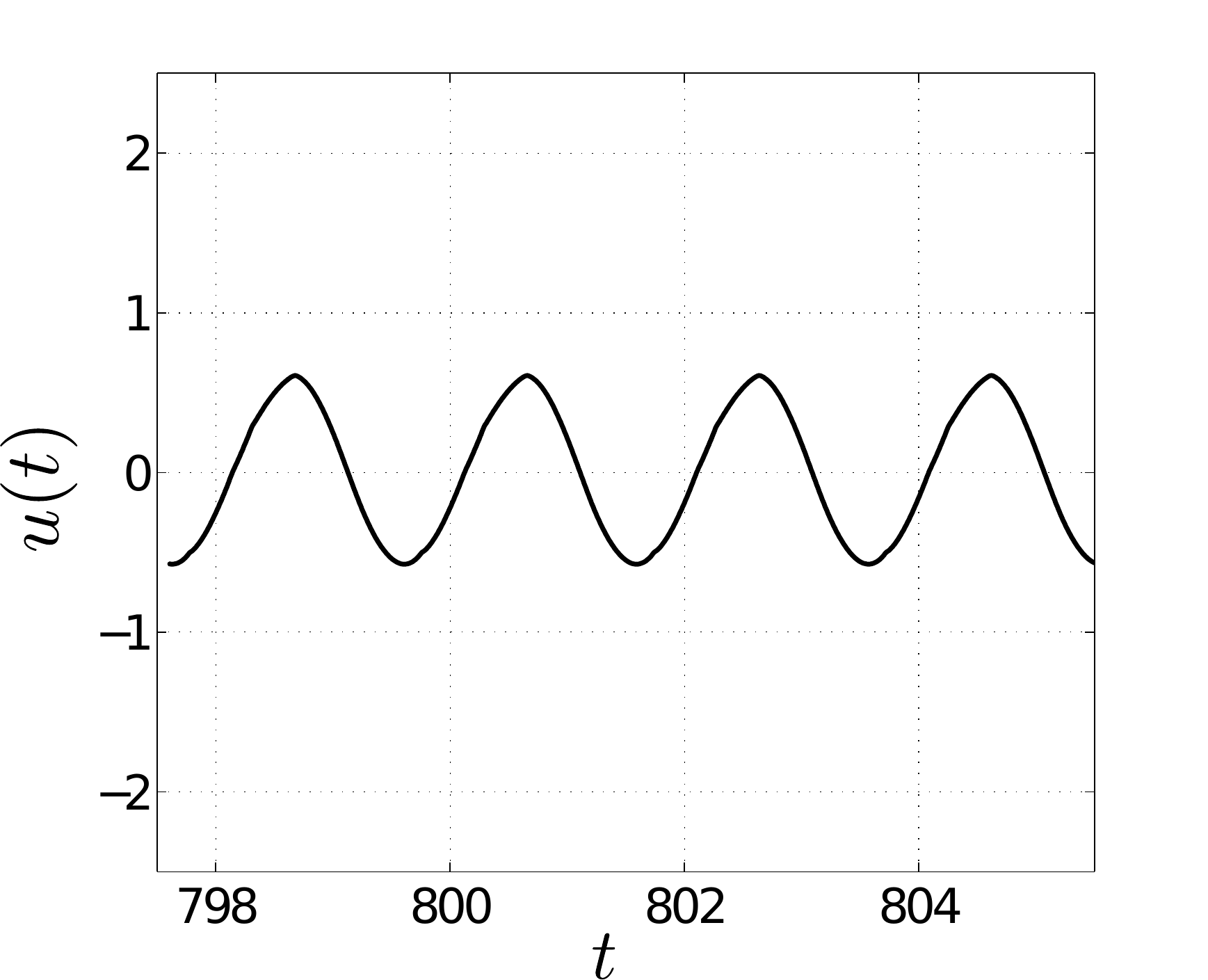}}
  \subfloat[$K = 1.0$]{\includegraphics[width = 0.26\textwidth] {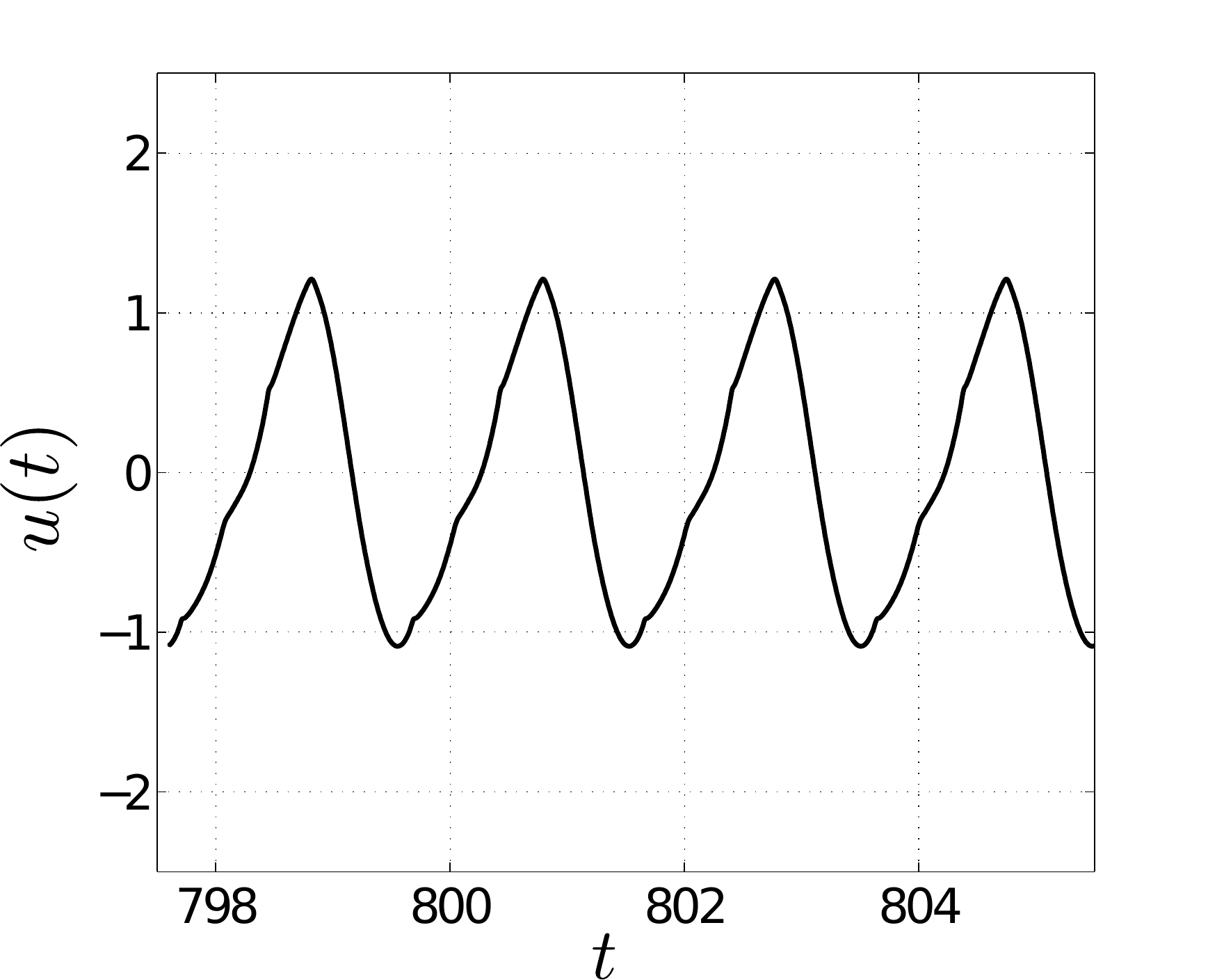}}
  \subfloat[$K = 1.5$]{\includegraphics[width = 0.26\textwidth] {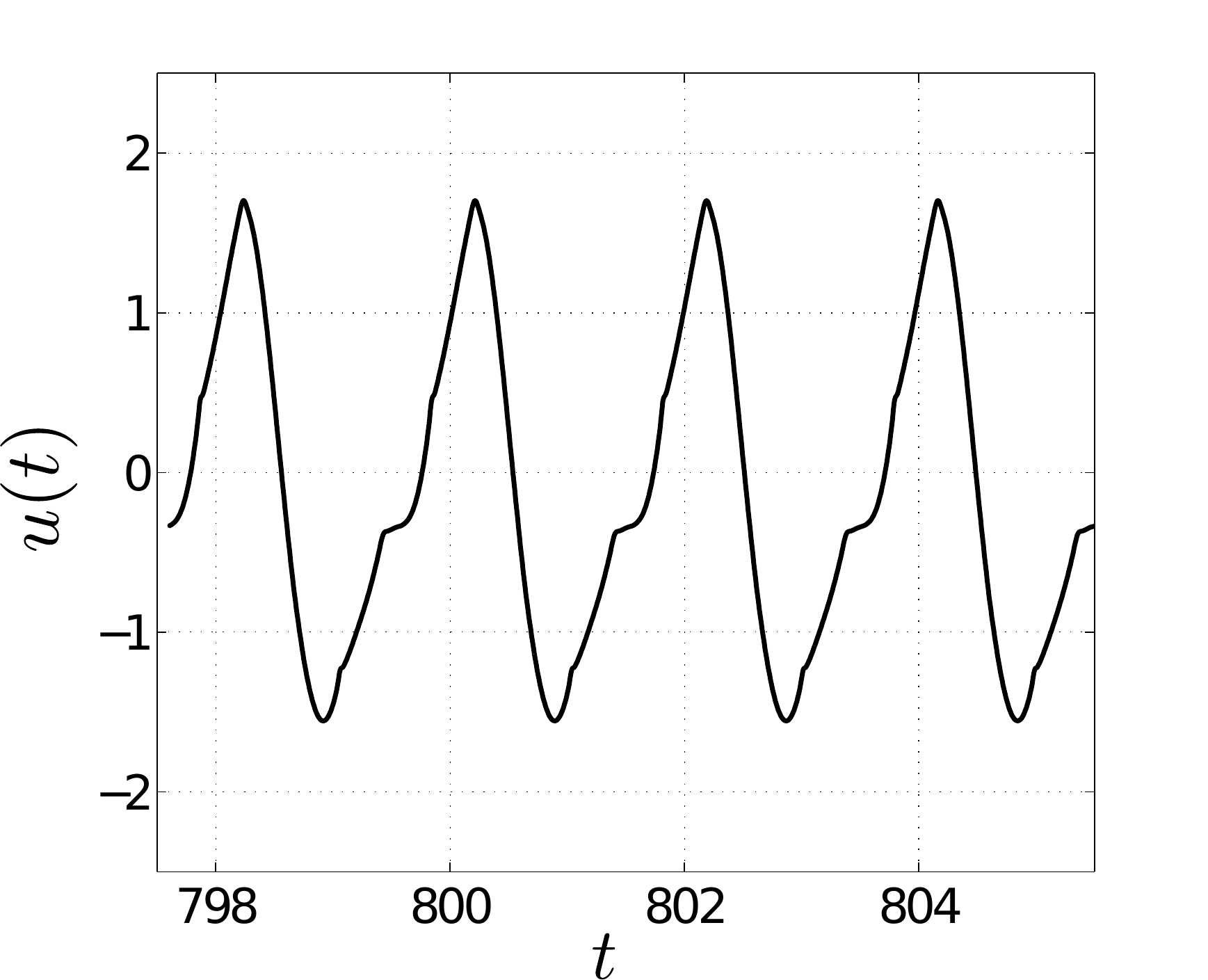}}
  \subfloat[$K = 1.9$]{\includegraphics[width = 0.26\textwidth] {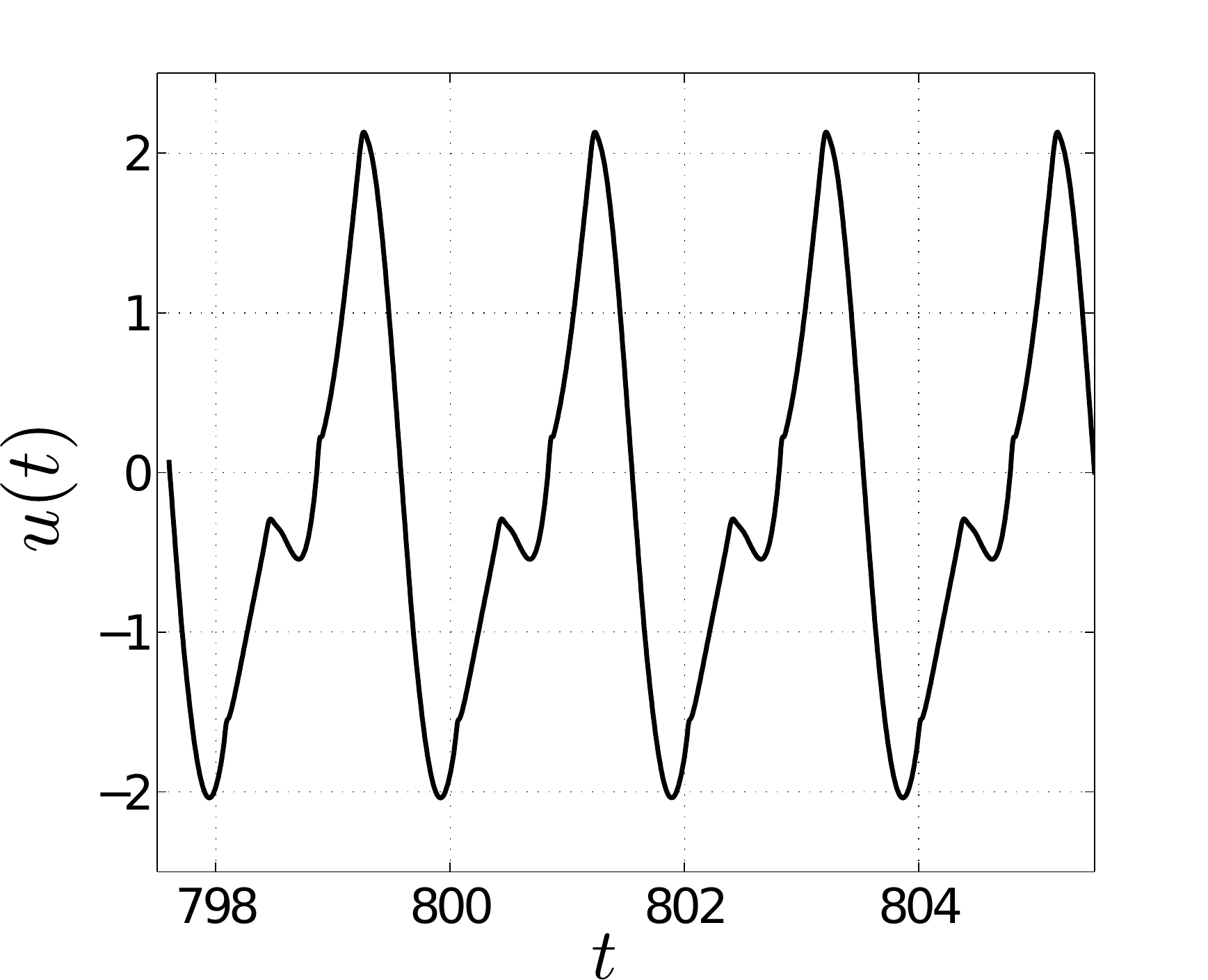}}\\
  \caption{Time series of the velocity signal at the probe location, $x = L/8$.}
  \label{fig:DMD_signal}
\end{figure}

Figure~\ref{fig:DMD_spec} shows the DMD-amplitudes of the selected limit cycles. The velocity signals corresponding to these spectra are also plotted in figure~\ref{fig:DMD_signal}. Similar to the eigenvalue spectra of the linear analysis, the modes appear in pairs. The mode with zero imaginary part is a stationary mode of the signal and approximates the mean flow.The spatial shape of the two pairs of the lowest frequency modes (together with the stationary mode) is shown in figure~\ref{fig:DMD_mode}. The stationary mode, ($\hbox{imag}(\lambda) = 0$), is in fact the discontinuous steady state, as opposed to the zero steady state of the linear regime, while the higher harmonics correspond to the eigenvectors with similar frequency and wavelength extracted by the linear analysis. Although the non-stationary modes are still discontinuous, the discontinuity is of much smaller relative amplitude, compared to the eigenvectors of the linear regime. This is not surprising as the total amplitude of the pressure and velocity signals has increased by orders of magnitude compared to the initial stage of linear growth (figure~\ref{fig:signal}). This finding suggests that even in the nonlinear regime, where the magnitude of the  discontinuity is much smaller than the amplitude of the signal, the acoustic modes are still affected by the presence of the heater, and the mean flow is strongly modified as well. This is not accounted for by any model using the Galerkin expansion, described in \S~\ref{sec:models}. The evolution of the DMD spectra shows that, as expected for the lower values of the heater strength, one mode is dominant throughout the saturated signal. The frequency of this mode is the same as the lowest acoustic frequency of the duct, and the shape of this mode is  shown in figure~\ref{fig:DMD_mode}. This suggest that for lower values of $K$ the first acoustic mode of the duct with $\hbox{imag}(\lambda) = \pi$, which is linearly unstable remains the only dominant mode in the nonlinear limit cycle. However, as the strength of the heater is increased the second harmonic of the first unstable mode becomes dynamically important in the saturated limit cycle and finally, at $K = 1.9$, reduces higher amplitudes in the DMD spectra than its lower harmonic. As a result, through nonlinear interactions the mode with frequency $\hbox{imag}(\lambda) = 2\pi$, which is linearly stable for this range of parameters becomes the dominant mode in the nonlinear limit cycle. The appearance of the second frequency can be identified by looking at the time series of the velocity signals. This frequency switching is not captured, by using the decoupled Galerkin approach. In conclusion, the choice of the numerical method affects the manner in which the signal is saturated through nonlinear interactions of different frequency modes.

\section{Summary and conclusions}
\label{sec:conclusion}

A dedicated high-fidelity solver designed for capturing the acoustic response of a system to a compact heat source has been designed and validated using the analytical solution of a simplified system. The eigenvalue spectrum and the neutral stability curves in the low-amplitude regime have been extracted using a collocation method designed for differential equations with delay. The eigenvalue spectra from this high-fidelity approach are compared to those of the analytic solution with no damping and show satisfactory agreement. The performance of the method in this linear regime is compared to models based on a Galerkin expansion using the continuous acoustic modes of the chamber in the absence of the heater. Our analysis shows that, due to the presence of a discontinuity in the velocity signal at the location of the heater, high-frequency oscillations appear in the eigenvectors computed by the Galerkin approach. Consequently, for damped systems which are designed to reproduce the experimental conditions, the eigenvalues of higher-frequency modes are damped compared to the high-fidelity solution, which also affects the marginal stability boundaries (neutral curves) of these modes. Analyzing the nonlinear evolution of the signal shows that depending on the value of heat release, $K$, the shape of the limit cycle varies from unimodal to bimodal. For low values of $K$, the limit cycle is dominated by a single frequency which is the unstable frequency of the linear stability diagram. At this stage, the evolution of the signal is similar between the decoupled Galerkin and the high-fidelity models. However, as $K$ increases the high-fidelity approach shows the appearance of a second frequency, which is the first harmonic of the unstable frequency.  The linear stability diagram confirms that this frequency is linearly stable for the range of parameters considered in this study but, due to nonlinearities, becomes significant as the limit cycle is established. Finally, as the value of $K$ is further increased, the higher harmonic becomes the dominant mode of the saturated limit cycle. In contrast to the high-fidelity approach, the decoupled Galerkin model predicts a unimodal limit cycle for a similar range of parameter space, as it does not account for the coupling between the modes. Although the coupled Galerkin approach predicts the appearance of the higher frequencies, it underpredicts the amplitude of the saturated limit cycle, since the shapes of the modes are not captured correctly by this approach. 

The appeal and use of Galerkin techniques to model thermoacoustic phenomena is largely driven by their simplicity in application and resulting equations. However, this ease of use comes at the expense of combining modeling assumptions with numerical approximations, hence making it more difficult to distinguish these two different error sources in the resulting behavior. For this reason, special care has to be exercised when choosing Galerkin techniques to discretize thermoacoustic equations that contain discontinuous terms.

The complex behavior studied here is accurately captured by the high-fidelity approach since its predictions are unaffected by the numerical scheme. This approach then allows the separation of numerical discretization and physical modeling assumptions, thus making possible a direct analysis  and comparison of various models related to heat release and damping.

\vspace{0.25in} This research was partially supported by the LASIPS Labex  from the Idex Paris-Saclay through the PROCID project , and by a fundamental project grant from ANR (French National Research Agency - ANR Blancs): Sechelles. The authors also wish to express their gratitude to Prof. Fred Culick for his interest in this work and for enlightening discussions on the Rijke-tube problem. Dr. Daniel Durox, Prof. Thierry Schuller and Luca Magri are gratefully acknowledged for fruitful discussions. 

\bibliographystyle{jfm}
\bibliography{paper}

\end{document}